\documentclass[aps,prb,amsmath,twocolumn,amssymb,floatfixng,showpacs,
superscriptaddress,footinbib]{revtex4-1}
\pdfoutput=1
\usepackage[dvips]{graphics}
\usepackage{bm}
\usepackage{epsfig}
\usepackage{enumerate}
\usepackage{subfigure}
\usepackage{color}
\usepackage{graphicx}
\usepackage{hyperref}
\hypersetup{
    pdfnewwindow=true,      
    colorlinks=true,       
    linkcolor=magenta,          
    citecolor=blue,        
    filecolor=blue,      
    urlcolor=blue        
}

\newcommand{\ie}{{\it i.e., }}

\newcommand{\bea}{\begin{eqnarray}}
\newcommand{\eea}{\end{eqnarray}}
\newcommand{\beq}{\begin{equation}}  
\newcommand{\eeq}{\end{equation}}

\begin{document} 
\title{Enhancement of thermoelectric performance of a nanoribbon made of $\alpha-\mathcal{T}_3$ lattice} 

\author{Mir-Waqas Alam}
\email{wmir@kfu.edu.sa}
\affiliation{Department of Physics, College of Science, King Faisal University, Al-Hassa 31982,
P.O. Box 400, Saudi Arabia}

\author{Basma Souayeh}
\email{bsouayeh@kfu.edu.sa}
\affiliation{Department of Physics, College of Science, King Faisal University, Al-Hassa 31982,
P.O. Box 400, Saudi Arabia}

\author{SK Firoz Islam}
\email{rafian.firoz@gmail.com}
\affiliation{Institute of Physics, Sachivalaya Marg, Bhubaneswar-751005, India}

\begin{abstract}
We present electronic and transport properties of a zigzag nanoribbon made of $\alpha-\mathcal{T}_3$
lattice. Our particular focus is on the effects of the continuous evolution of the edge modes (
from flat to dispersive) on the thermoelectric transport properties. Unlike the case of graphene
nanoribbon, the zigzag nanoribbon of $\alpha-\mathcal{T}_3$ lattice can host a pair of 
dispersive (chiral) edge modes at the two valleys for specific width of the ribbon.
Moreover, gap opening can also occur at the two valleys depending on the width. The slope
of the chiral edge modes and the energy gap strongly depend on the relative strength of two
kinds of hoping parameters present in the system. We compute corresponding transport
coefficients such as conductance, thermopower, thermal conductance and the thermoelectric
figure of merits by using the tight-binding Green function formalism, in order to explore
the roles of the dispersive edge modes. It is found that the thermopower and thermoelectric
figure of merits can be enhanced significantly by suitably controlling the edge modes.
The figure of merits can be enhanced by thirty times under suitable parameter regime in
comparison to the case of graphene. Finally, we reveal that the presence of line defect, 
close to the edge, can cause a significant impact on the edge modes as well as on electrical
conductance and thermopower.
\end{abstract}

\maketitle
\section{Introduction}
The discovery of graphene~\cite{novo,neto} has boosted the search of graphene-like two-dimensional
Dirac materials because of their peculiar band structure and possible technological
applications. The electronic properties of Dirac materials are described by the linear band
dispersion in low energy regime. The $\mathcal{T}_3$ or dice lattice~\cite{vidal}
is the graphene-like $2$D material with an additional atom at the centre of hexagon.
One of the  unique feature of such material is that its quasi particles exhibit higher
pseudo spin $S=1$ states~\cite{vidal} unlike $1/2$ in graphene. Apart from it, the additional
atom in the $T_3$ lattice causes dispersionless flat band at each valley in addition
to the Dirac cones~\cite{vidal}. In recent times, much attention have been paid on such
Dirac-Weyl materials with higher spin states, $S=1$, $3/2$,$2$,etc.~\cite{malcolm,janik,
lan,morigi}, in order to reveal the roles of the additional atom.

The $T_3$ lattice (pseudo spin $S=1$) can be smoothly interpolated to the graphene (pseudo spin $S=1/2$) 
by using the $\alpha$-$\mathcal{T}_3$ model. Here, $\alpha$ is related to the strength of the
hoping between the central atom to its nearest neighbors and ranges from `$0$' (graphene) to
`$1$' (dice lattice or $T_3$) lattice. It has recently been shown in Hg$_{1-x}$Cd$_{x}$Te
that this material can be mapped to $\alpha$-$\mathcal{T}_3$ model~\cite{malcolm} with $\alpha=1/\sqrt{3}$ under
the suitable doping concentration. The continuous evolution from graphene to dice lattice
by using $\alpha-\mathcal{T}_3$ model has been extensively exploited in unusual Hall
conductivity~\cite{nicol1,tutul1}, Weiss oscillation\cite{firoz_para}, Klein tunneling\cite{daniel,klein},
optical~\cite{malcolm,nicol3,dora3,plasmon,newT3} properties, irradiation effects\cite{photo},
topological properties\cite{dora2} and wave packet dynamics\cite{tutul2}.
However, most of the study of transport properties in the $\alpha$-$\mathcal{T}_3$ lattice are
limited to the bulk in spite of the fact that electronic band structure as well as the
transport phenomena are very sensitive to the edge geometry of honeycomb lattice\cite{neto,newT3}.

The thermoelectric properties of material\cite{nolas} have been always under active consideration
among research community for its ability to probe the electronic system and potential technological
applications\cite{disalvo,snyder}. The thermal gradient across the two ends of an electronic
system can drive charge carriers from hotter to cooler end and can generate a voltage gradient
across these two ends-known as thermopower (S) per unit temperature gradient. Apart from the 
thermal transport in the bulk of $2$D hexagonal lattice\cite{das_sarma,nam,hao,yuri,wei}, several
works have demonstrated that thermoelectric performance can be further improved by considering nanoribbon
of graphene\cite{hossain,van} or black phosphorus\cite{Ma,flores}.

In this work, we first address the energy band dispersion of the nanoribbon of such material
by using tight binding method. Here, we particularly focus on zigzag edge only as it hosts a pair of edge modes.
We observe that unlike the case of graphene, the zigzag nanoribbon of $\alpha-\mathcal{T}_3$ exhibits
gapless dispersive edge modes (chiral edge modes) for width of $N=3q+1$ ( $q$ is the positive integer).
On the other hand, the edge modes are gapped for the width $N\ne 3q+1$. This is in contrast to
the case of a zigzag ribbon of graphene\cite{neto} where edge modes are dispersionless, gapless
and not chiral. Subsequently we use tight-binding Green function approach to obtain the conductance,
thermopower and thermoelectric figure of merits of such ribbon. We found that thermopower and
figure of merits can be enhanced significantly by controlling the features of edge modes by means of $\alpha$. 
Finally, we discuss the effects of line defects on edge modes and transport properties.

The paper is organized as follows: In Sec.~\ref{sec2}, we discuss the tight-binding Hamiltonian
and energy band dispersion for zigzag nanoribbon. A brief review of the tight-binding Green function
formalism for the evaluation of transport coefficients are presented in Sec.~\ref{sec3}.
In Sec.~\ref{sec4} we present our numerical results and discussions.
Finally, we summarize our results and conclude in Sec.~\ref{sec5}.
\section{Tight binding Hamiltonian and energy dispersion}\label{sec2}
\begin{figure}[!thpb]
\centering
\includegraphics[height=3cm,width=0.50 \linewidth]{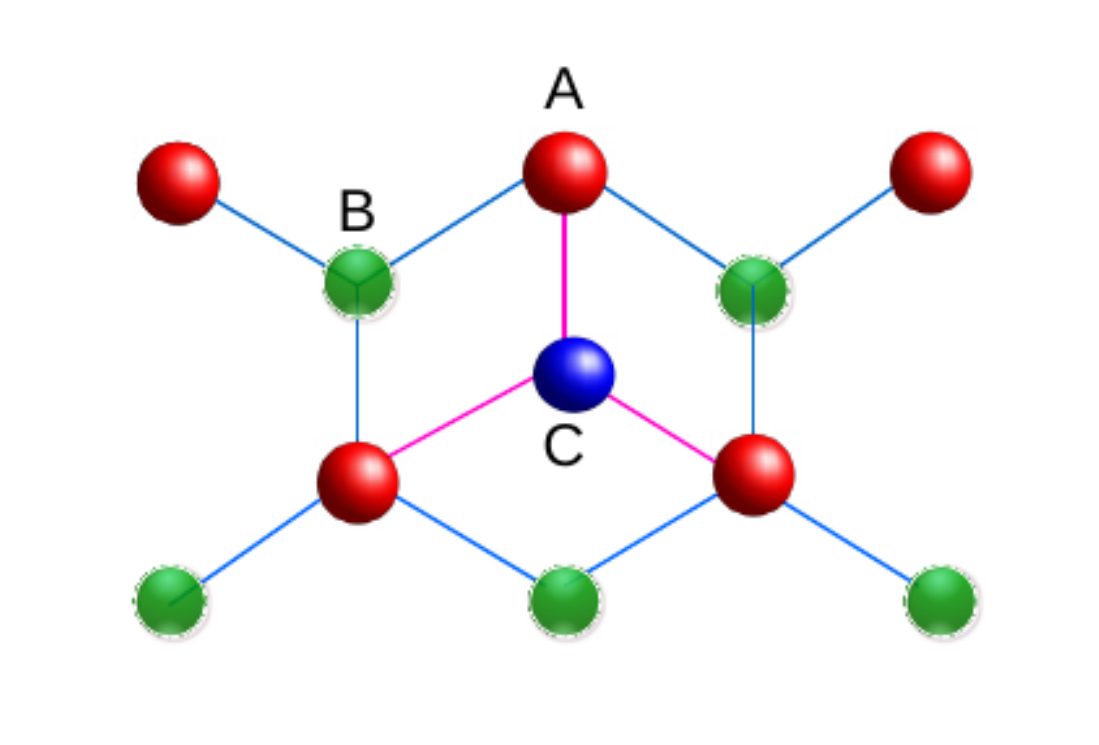}
\caption{(Color online) Schematic of the hexagon of $\alpha$-$\mathcal{T}_3$ lattice. Three different colors 
are used to denote denote three sub-lattices, \ie A (red), B (green) and C (blue). The two different
hoping terms $t$ and $\alpha t$ are denoted by blue and magenta lines, respectively.}
\label{model}
\end{figure}

 In this section, we first present a brief description of the lattice geometry of the alpha-$\mathcal{T}_3$ lattice.
 This lattice mimics the geometry of graphene monolayer with an additional atom at the centre of the 
 hexagon. A typical sketch of its hexagon is shown in the Fig~(\ref{model}). It has two different hoping 
 parameters. The hoping parameter between $A$ and $B$ sublattices is denoted by `$t$'
 whereas $\alpha t$ is between the subalttice $C$ and $B$. 
The tight-binding Hamiltonian of this lattice, without any spin-orbit coupling is given by
\begin{equation}
 H_0=\sum_{<ij>}t_{ij}c_{i}^{\dagger}c_j+\sum_{<i,l>}t_{i,l}c_{i}^{\dagger}c_{l}+h.c \ ,
\label{H}
\end{equation}
where the summation index $i$, $j$ and $l$ run over $A$, $B$ and $C$ sublattices. 
The relevant hoping parameters are $t_{i,j}=t$ and $t_{i,l}=\alpha t$. 
The creation (annihilation) operators at $i$-th site are denoted by $c_{i}^{\dagger}$ ($c_{i}$).
\begin{figure}[!thpb]
\centering
\includegraphics[height=4.8cm,width=0.58 \linewidth]{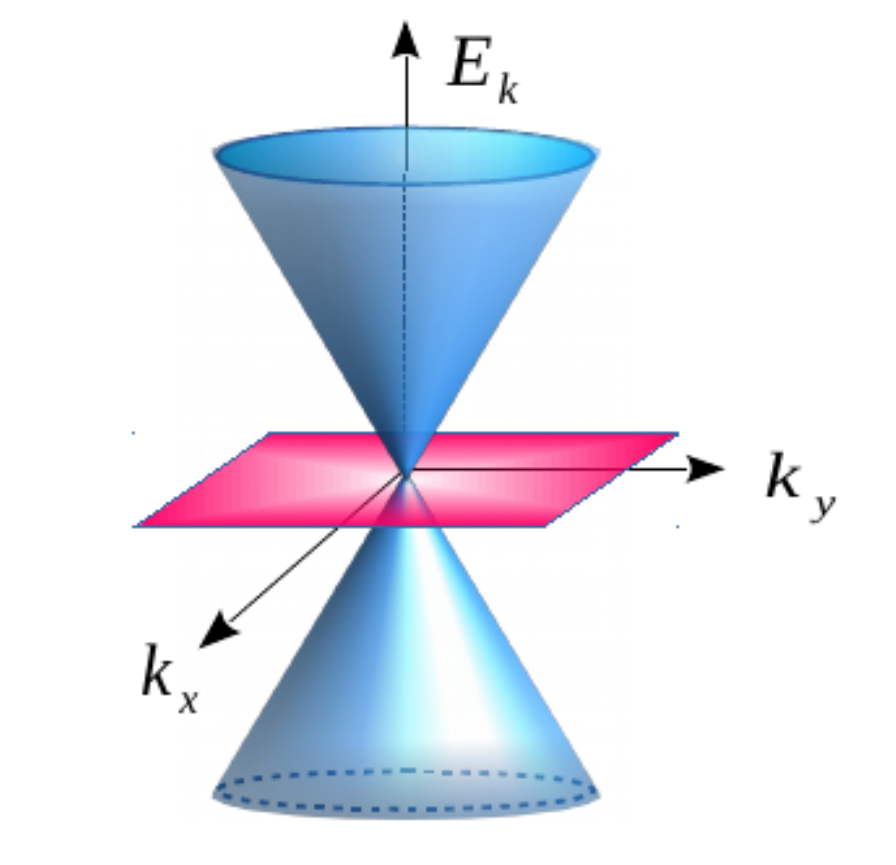}
\caption{(Color online) Schematic of the band structure of $\rm \alpha$-$\mathcal{T}_3$ lattice. The blue colored
conic bands represent graphene-like linear dispersion whereas the dispersionless flat band is distinguished by
pink color.}
\label{band}
\end{figure}

However, we briefly comment here that the Hamiltonian in the continuum model inside the bulk without
any boundary can be written in three sublattice space as\cite{morigi}
\begin{equation}
\mathcal{H}_0=\left[\begin{array}[c]{ccc}
            0      &  g_{\bf p}\cos\phi &   0 \\
            g_{\bf p}^\ast\cos\phi  &  0  &  g_{\bf p}\sin\phi \\
            0     &  g_{\bf p}^\ast\sin\phi  &  0\end{array}\right].
\end{equation}
Here, $g_{\bf p}=v_F(\xi p_x-ip_y)$ where $\xi=\pm$ denotes the two valleys $K$ and $K^{\prime}$, respectively. 
${\bf p}=\{p_x,p_y\}$ is the $2$D momentum vector and $v_F$ is the Fermi velocity.
Note that, the angle $\phi$ is related to the $\alpha$ as $\phi=\tan^{-1}\alpha$. The energy dispersion 
of the above Hamiltonian is linear as $E_{k,\lambda}=\lambda \hbar v_{F}k$, with $\lambda=\pm$ 
correspond to band index, as shown in Fig.~(\ref{band}).
\begin{figure}[!htb]
\centering
\includegraphics[width=.5\textwidth,height=5cm]{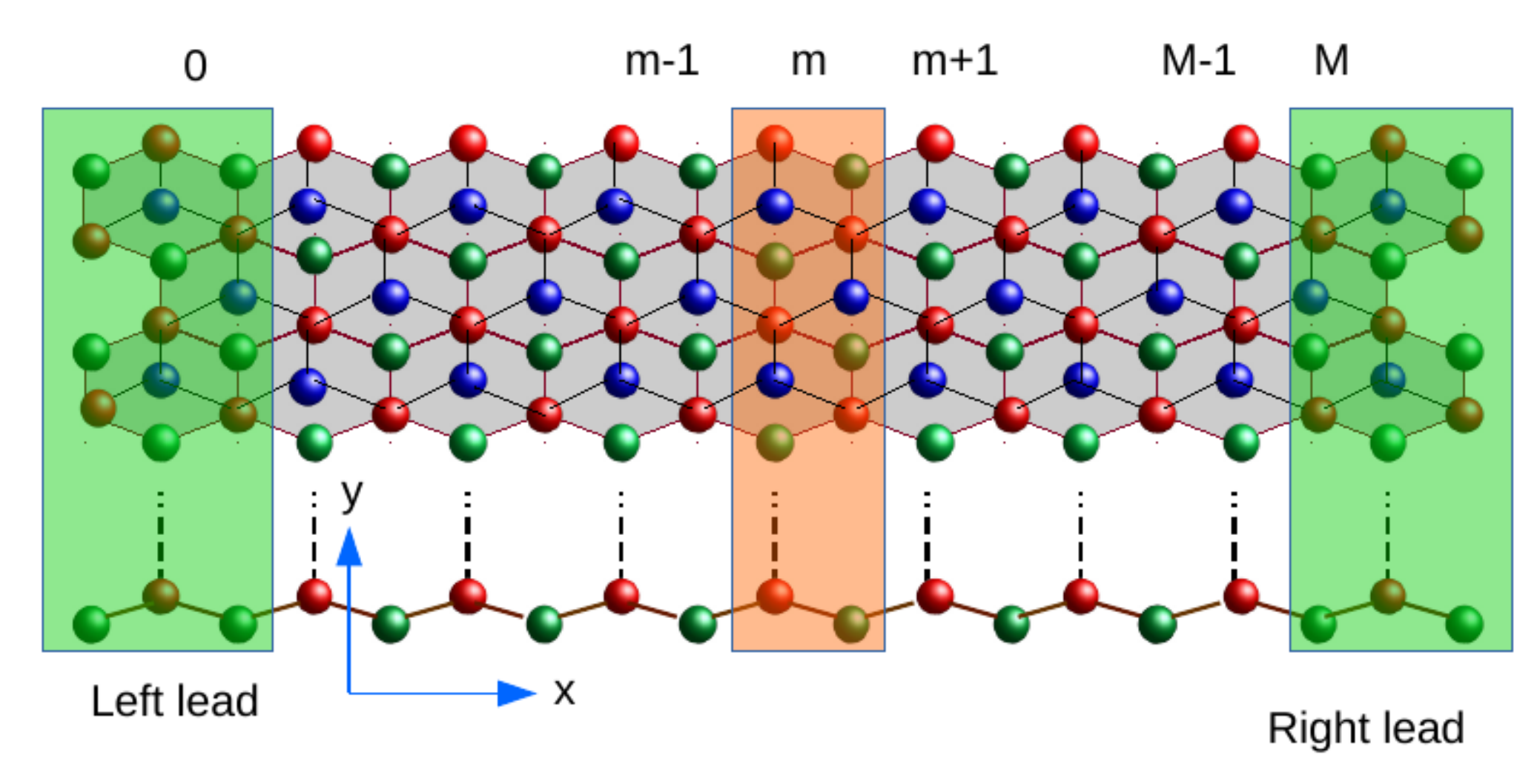}
\caption{Schematic of zigzag nanoribbon of $\alpha-T_3$ lattice. The unit cell is shown by
rectangular orange shadowed region and denoted by `m' index. The central region is attached
to the two leads at left and right end which are shown by light green shadowed region.}
\label{ribbon}
\end{figure}
It is also worthwhile to mention that the central atom does not play any role in 
the conic bands except the appearance of dispersionless flat band. However, in presence of 
magnetic field the C atoms (and hence $\alpha$) can lift valley degeneracy in the Landau levels\cite{nicol1,tutul1}
as well as can give rise to the unusual Hall conductivity. This is in contrast to the graphene
where the Landau levels at two valleys are identical (degenerate).

In the present study, we particularly focus on such lattice with finite width (nanoribbon),
which has not been considered previously in the context of transport.
\begin{figure}[!thpb]
\centering
\includegraphics[height=5cm,width=0.4\linewidth]{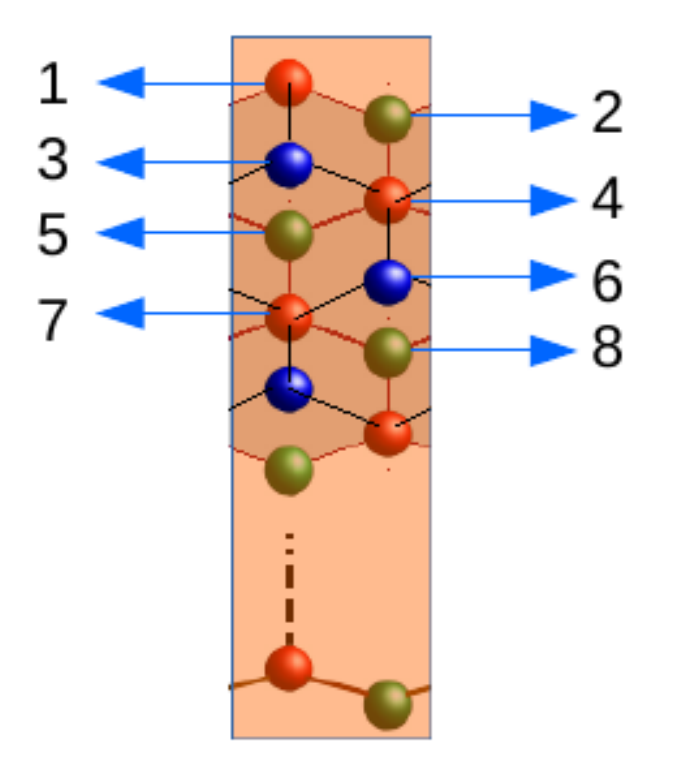}
\caption{The numbering of the atoms in each unit cell is shown explicitly.}
\label{number}
\end{figure}
 \begin{figure}[htb]
\begin{minipage}[t]{0.5\textwidth}
  \hspace{-.4cm}{ \includegraphics[width=.5\textwidth,height=4cm]{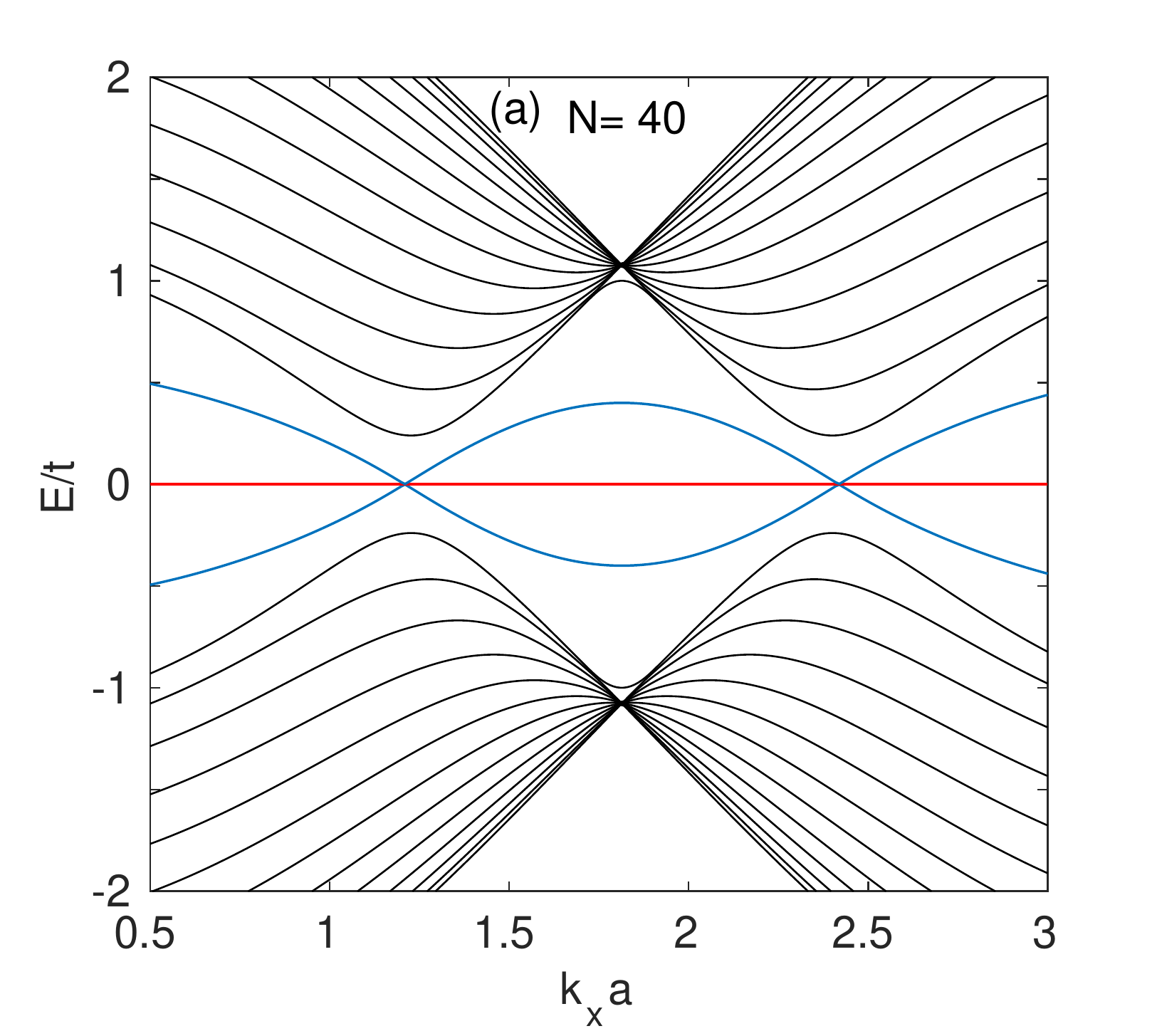}}
  \hspace{-.5cm}{ \includegraphics[width=.5\textwidth,height=4cm]{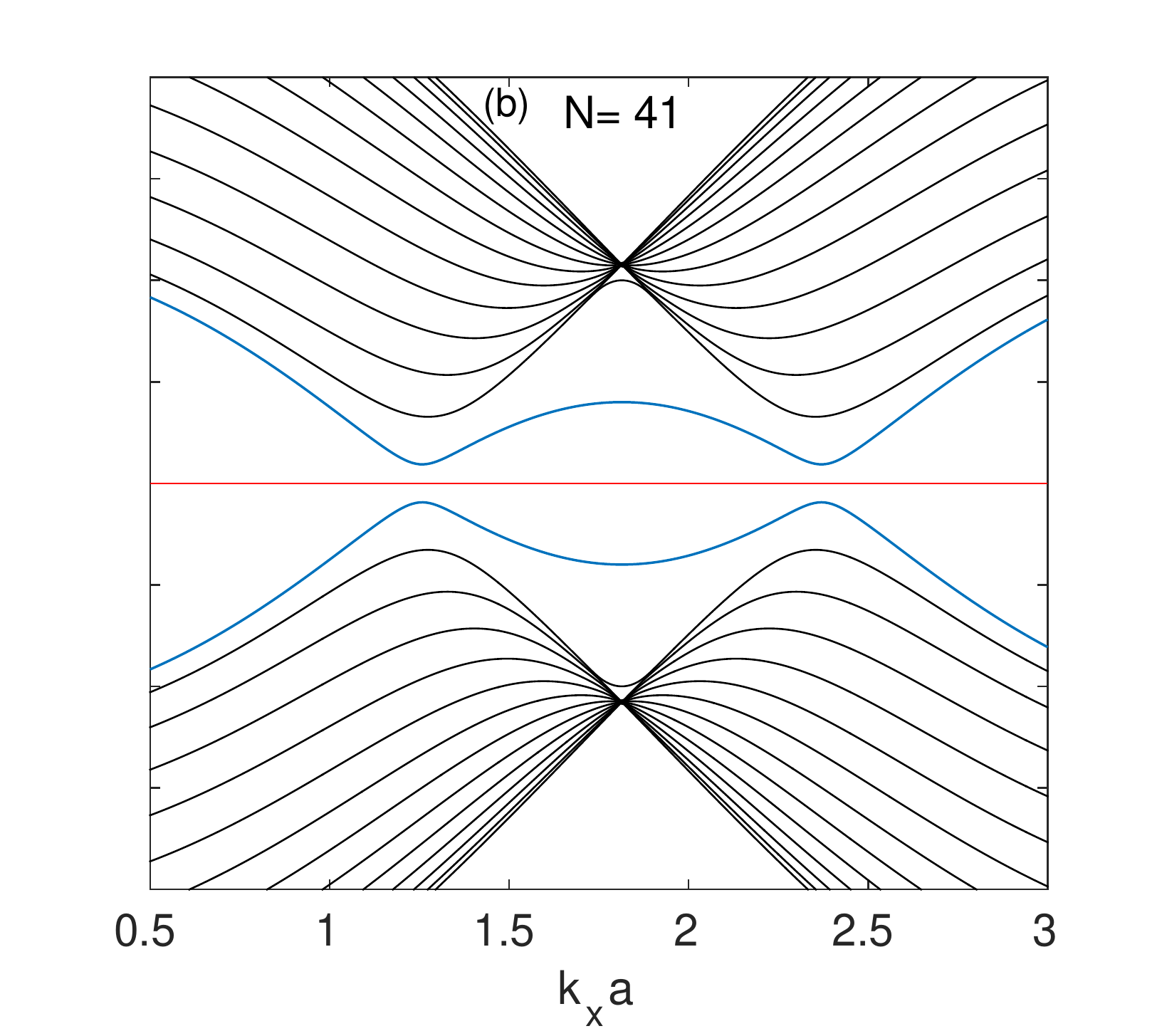}}
\end{minipage}
\begin{minipage}[t]{0.5\textwidth}
  \hspace{-.4cm}{ \includegraphics[width=.5\textwidth,height=4cm]{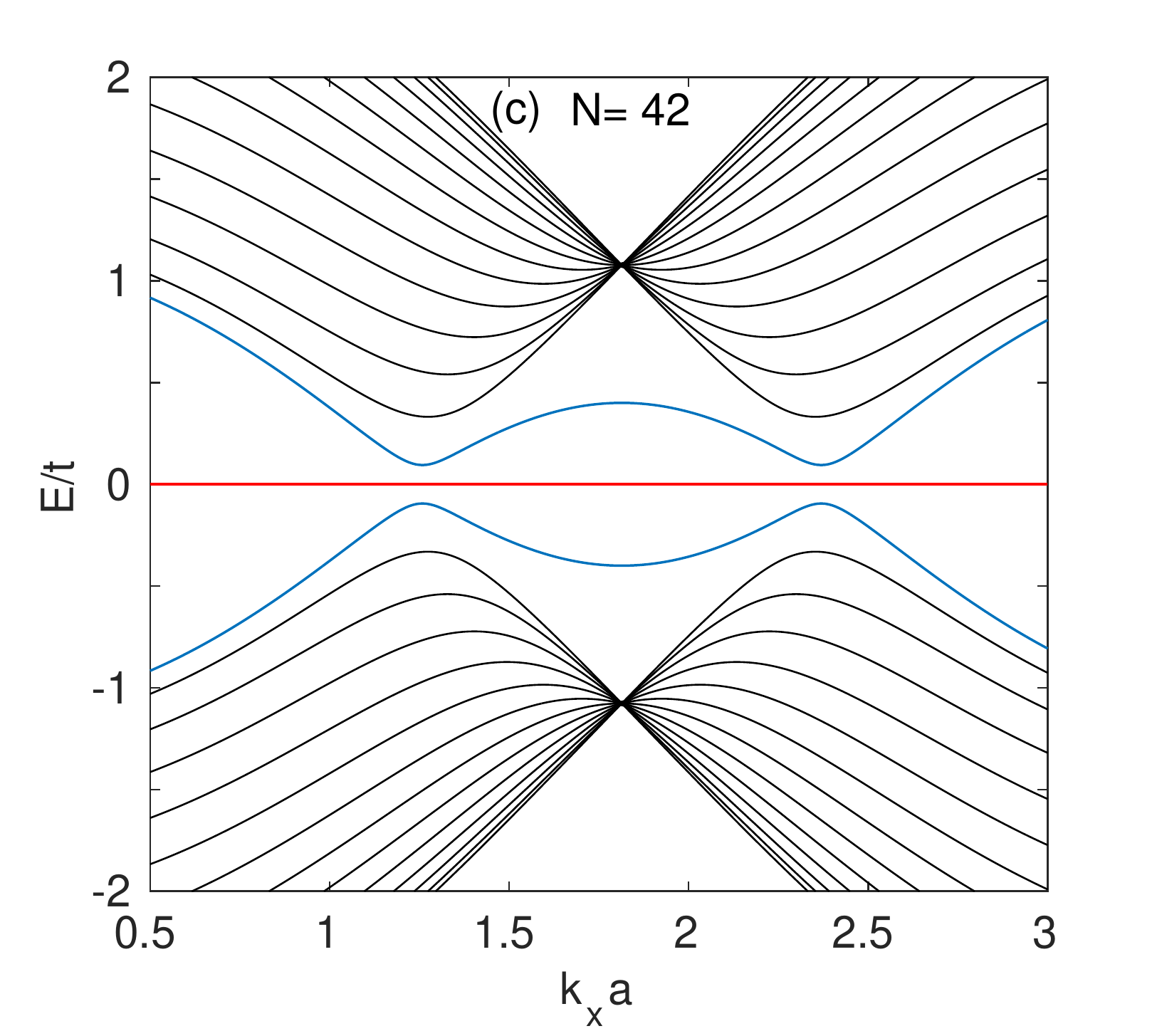}}
  \hspace{-.5cm}{ \includegraphics[width=.5\textwidth,height=4cm]{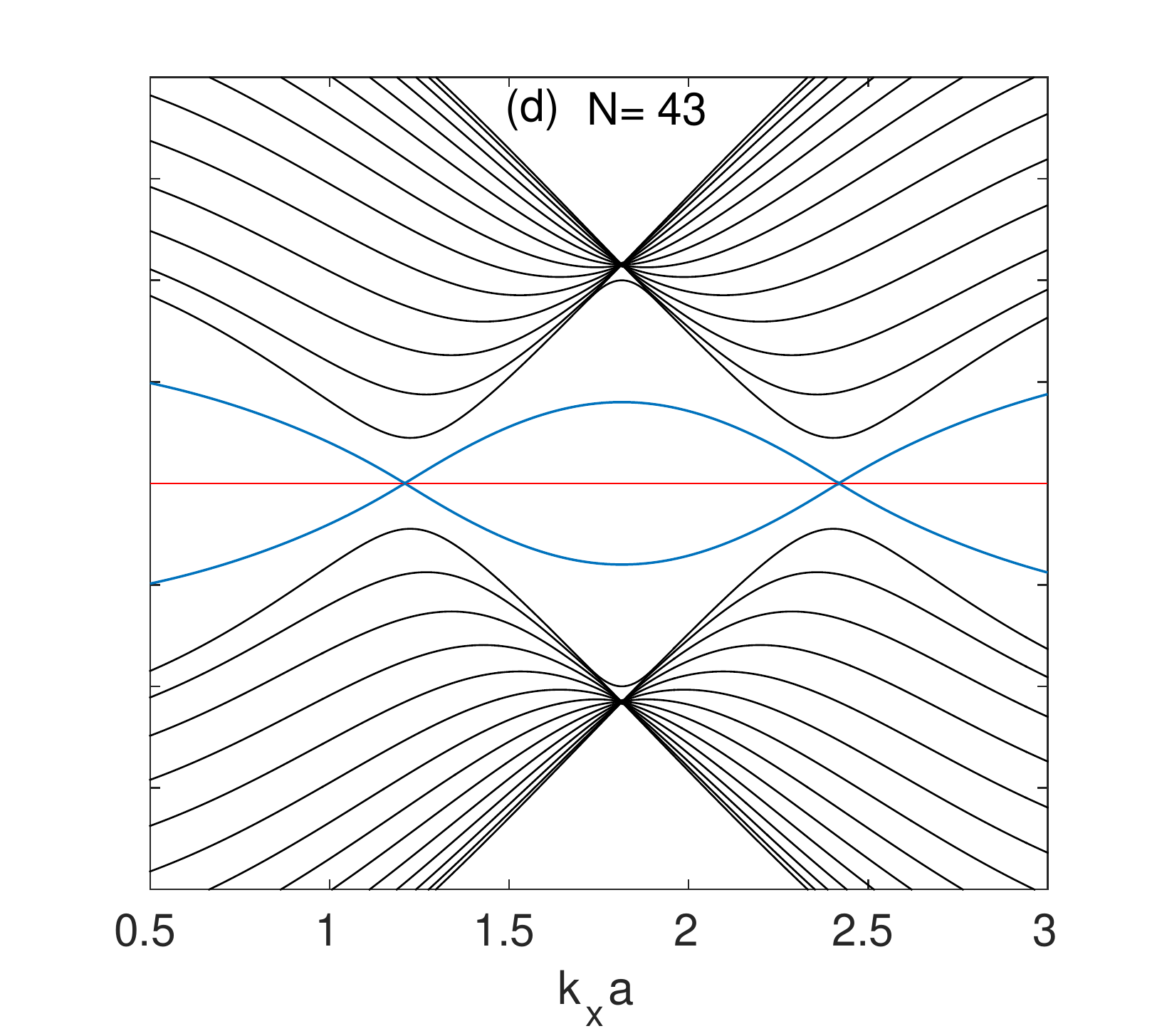}}
\end{minipage}
 \caption{Energy spectrum (in units of t) for zigzag edged nanoribbon for (a) $N=40$ (b) 
 $N=41$ (c) $N=42$ and (d) $N= 43$. We have taken
 $\alpha= 0.4$ for all of them.}
 \label{band40}
 \end{figure}
 The nanoribbon is considered to be infinitely extended along the $x$-direction with a finite
 width along the $y$-direction. The nanoribbon can be thought as a linear chain made of 
 iterative unit cells as shown by the rectangular shaped orange shadowed region in Fig.~\ref{ribbon}.
 The width of the nanoribbon is given by $N$-the number of atoms per unit cell. To study the 
 transport properties, we consider a two terminal device which consists of three regions as
 shown in the Fig.~(\ref{ribbon}). The central region is made of zigzag ribbon which is
 attached to the left and right identical leads. The locations of all the unit cells
 forming the left and right leads are at $-\infty,-1,0$ and $M,M+1...\infty$, respectively.
 Whereas the central regions are composed of the unit cells at $1,2,3...M$. By implementing
 Bloch's theorem, total Hamiltonian of the device can be written as
\begin{equation}
 \mathcal{H}_{k_x}=\mathcal{H}_{00}+\mathcal{H}_{-10}e^{-ik_xa}+\mathcal{H}_{01}e^{ik_xa},
 \end{equation}
where $\mathcal{H}_{00}$ is the on-site energy matrix of the unit cell at site $0$. On the other hand,
$\mathcal{H}_{01}$ or $\mathcal{H}_{-10}$ denotes the coupling matrix between the left and right
adjacent unit cells. Here, $a$ is the unit cell separation. The numbering of atoms in each unit cell
is shown in Fig.~(\ref{number}), in order to construct the Hamiltonian matrix. The momentum ($k_x$)
along the $x$-direction is conserved as the ribbon is translationally invariant along this direction. 

\begin{figure*}[!thpb]
\centering
\includegraphics[height=4.6cm,width=0.32\linewidth]{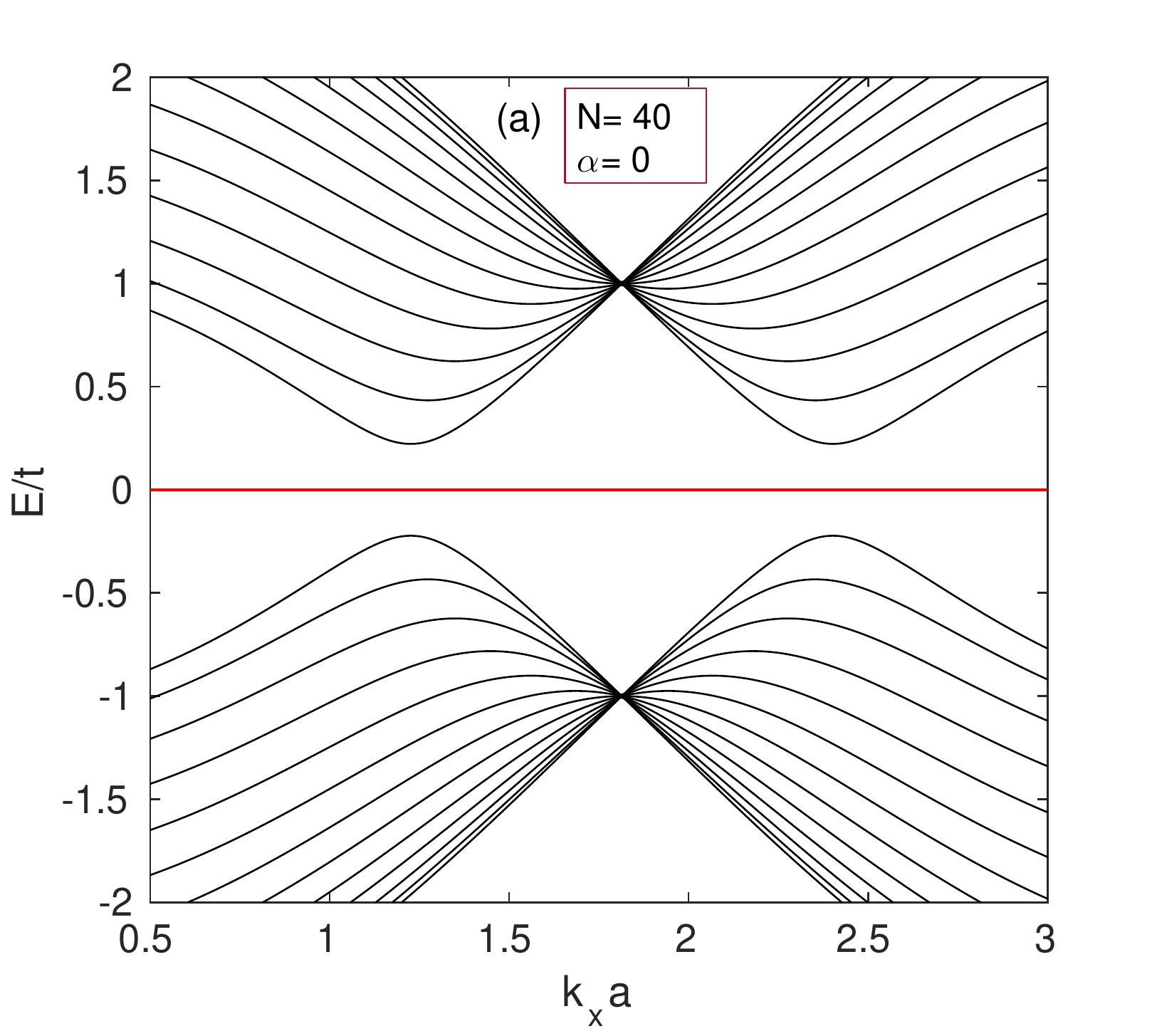}
\includegraphics[height=4.6cm,width=0.32\linewidth]{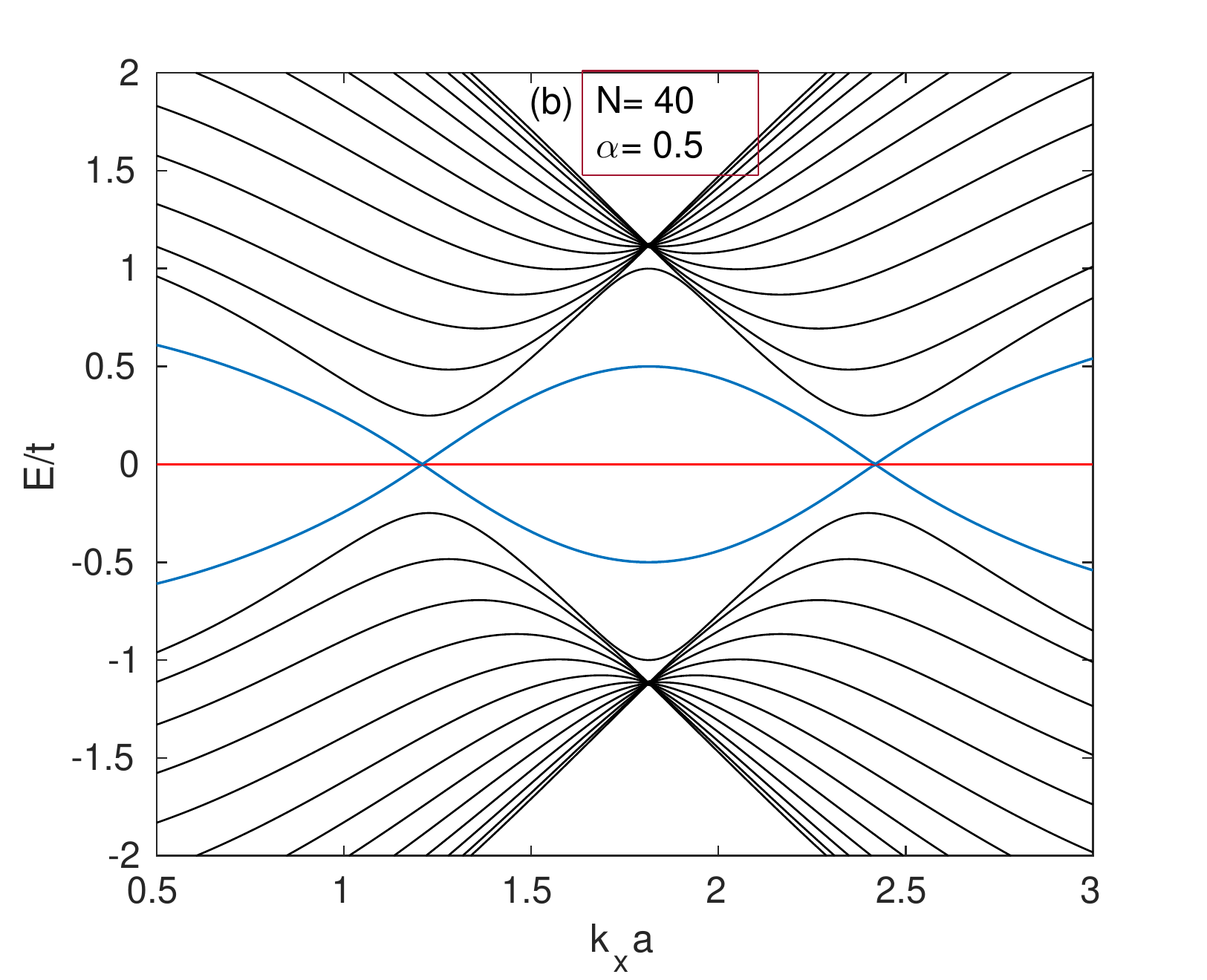}
\includegraphics[height=4.6cm,width=0.32\linewidth]{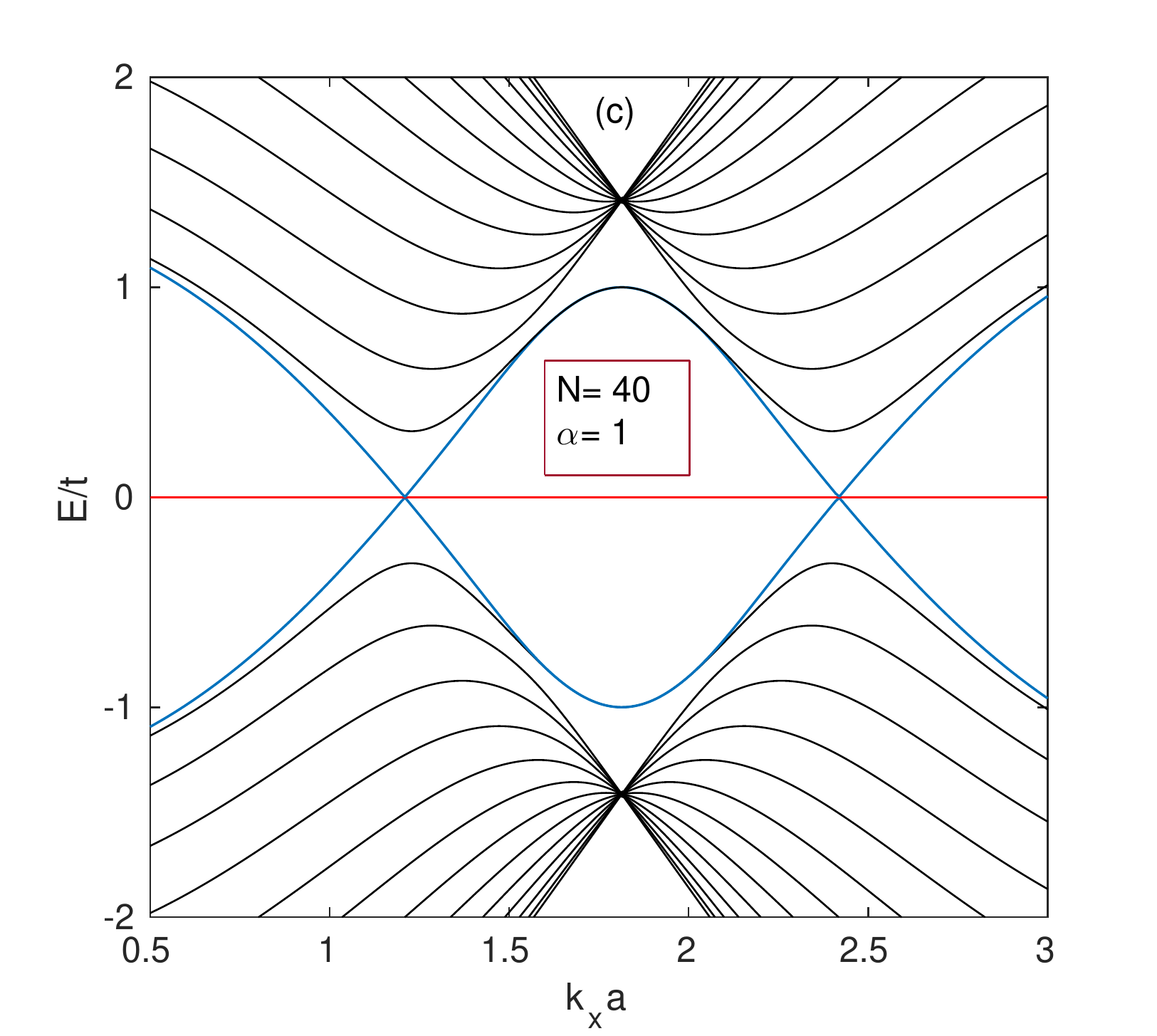}
\caption{Energy dispersion for (a) $\alpha=0$ (b) $\alpha=0.5$ and (c) $\alpha=1.0$. The width is taken as
$N=40$ i.e, gapless dispersion. The energy is taken in  units of t.}
\label{alpha_forN40}
\end{figure*}
\begin{figure*}[!thpb]
\centering
\includegraphics[height=4.5cm,width=0.32\linewidth]{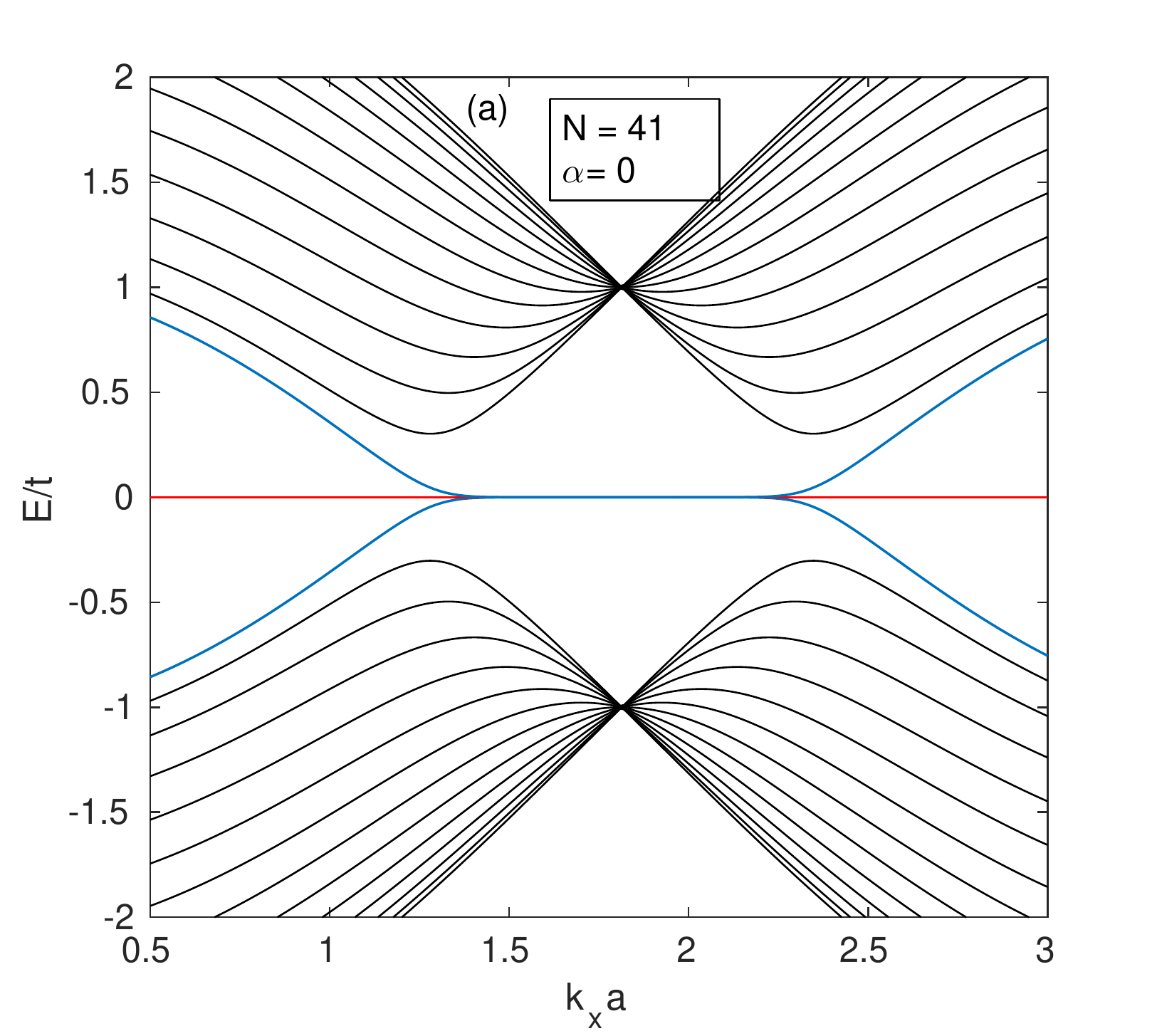}
\includegraphics[height=4.5cm,width=0.32\linewidth]{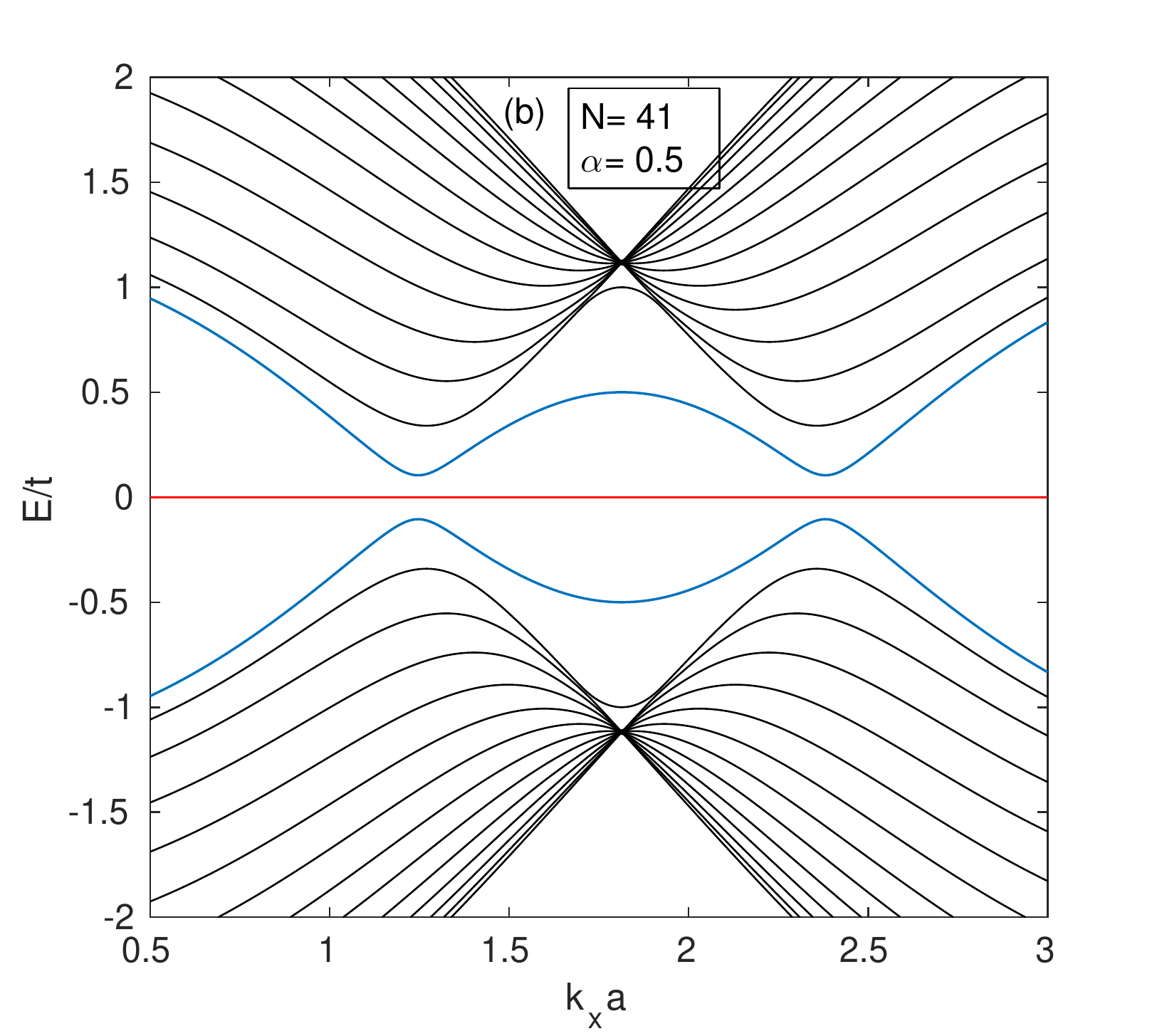}
\includegraphics[height=4.5cm,width=0.32\linewidth]{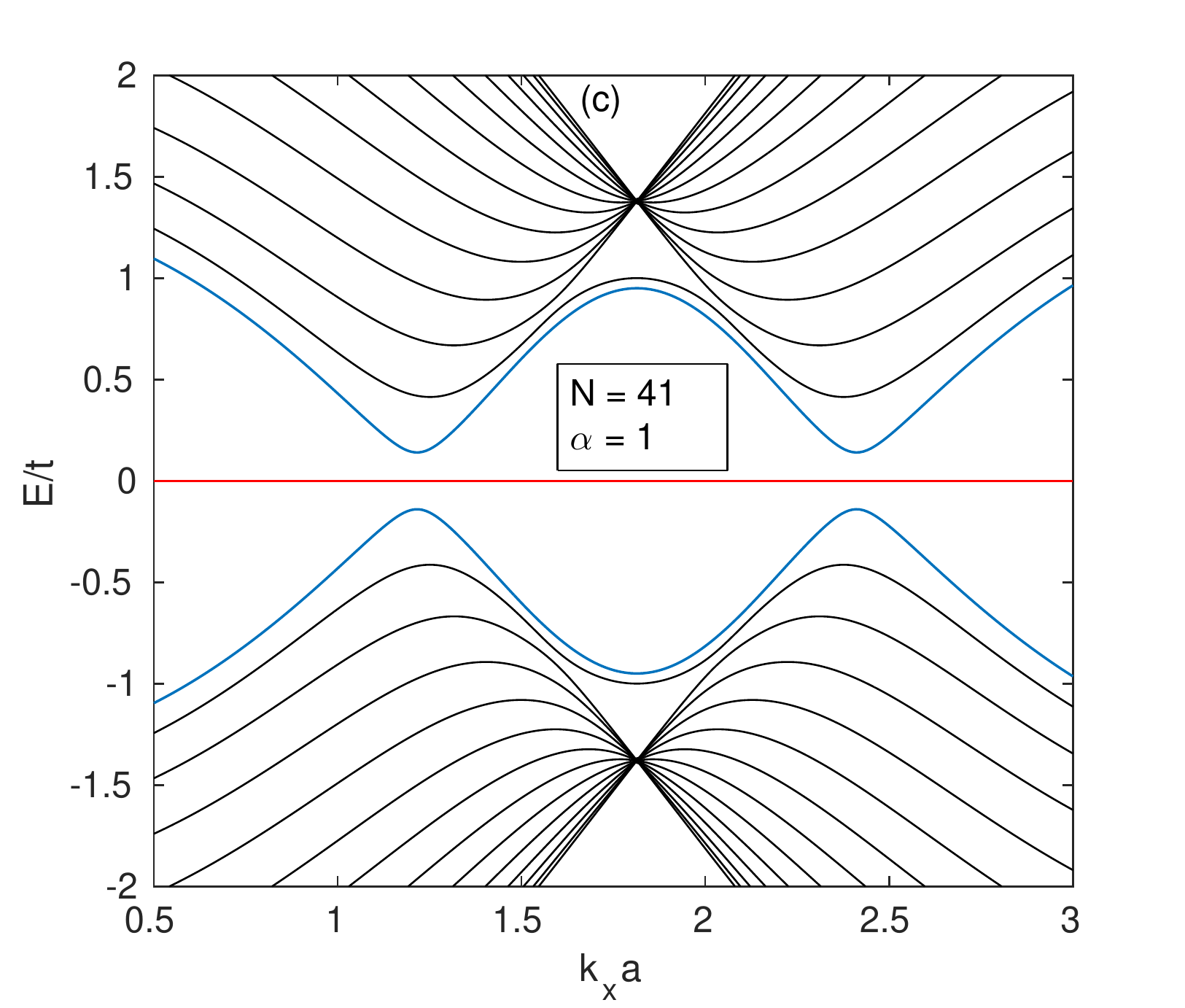}
\caption{Energy dispersion for (a) $\alpha=0$ (b) $\alpha=0.5$ and (c) $\alpha=1.0$. The width is taken as
$N=41$ i.e, gapped dispersion}
\label{alpha_forN41}
\end{figure*}
We solve numerically the above equation to obtain the energy dispersion of the nanoribbon
and plotted in Fig.~(\ref{band40}). We observe that edge modes (sky blue line) can be
gapless dispersive or gapped depending on the width. The gapless chiral edge modes appear for the width
$N=3q+1$, otherwise gapped. The hoping parameter between $B$ and $C$ sublattices are taken
corresponding to $\alpha=0.4$ in both cases. The Fig.~(\ref{band40})a and Fig.~(\ref{band40})d 
show gapless chiral edge modes for the widths $N=40$ and $43$ which satisfy the condition of 
width $N=3q+1$ and of course both edges are composed of $A$ and $B$ sublattices only.
On the other hand, Fig.~(\ref{band40})b and Fig.~(\ref{band40})c exhibit a pair of gapped
edge modes for widths $N=41$ and $42$ (i.e., when width $N\ne 3q+1$). Note that the crossing of the edge 
modes for gapless dispersion is the outcome of the additional hoping parameter due to the presence
of C sublattices in addition to the usual three nearest neighbor sublattices.

Now we examine how the variation of $\alpha$ affects the features
of chiral and gapped edge modes. First, we plot the energy dispersion for different values
of $\alpha$ in Fig.~(\ref{alpha_forN40}). The Fig.~(\ref{alpha_forN40})a is plotted for $\alpha=0$
and it enforces edge modes to collapse on the dispersionless flat band. Note that the 
width $N=40$ is corresponding to the case of non-identical edges for $\alpha=0$ i.e,
one edge is zigzag and another one is Klein-edge shape as named in Ref.~[\onlinecite{lakshmi}].
In such case a gap can be seen between flat band and other transverse modes which 
is also in agreement with the results based on the Harper equation in Ref.~[\onlinecite{newT3}]
and analytical work in Ref.~[\onlinecite{LTP}]. With the increase of $\alpha$,
edge modes emerges and exhibit dispersive feature
[see Fig.~(\ref{alpha_forN40})b] and the slope of which increases further with
$\alpha=1.0$ as shown in Fig.~(\ref{alpha_forN40})c. This slope actually gives rise to
the non-zero group velocity and hence induces significant contribution to the transport properties.

To recover the spectrum of zigzag nanoribbon of graphene, we  must make sure that both edges are 
composed of $A$ and $B$ sublattices which corresponds to the case of $\alpha=0$ and width $N=41$.
The Fig.~(\ref{alpha_forN41})a is plotted for these parameters, which shows the dispersionless
edge modes as in graphene nanoribbon except the presence of the flat band. Note that the flat band 
corresponds to the presence of $C$ sublattices. We plot the same for $\alpha=0.5$ in Fig.~(\ref{alpha_forN41})b
which shows that a gap opening occurs between the edge mode and the zero energy flat band. This 
gap opening increases slowly with the further increase of $\alpha$ as shown in Fig.~(\ref{alpha_forN41})c

Note that we are not considering the case of armchair edged ribbon as it does not give 
any new feature to the band dispersions in comparison to the armchair edged graphene\cite{neto,LTP}.

\section{Basic formalism of TB Green function approach}\label{sec3}
 In this section, we discuss the formalism to calculate different transport  coefficients under
 thermal/potential gradient. Let a temperature gradient of $\nabla T$ is applied between the
 left and right lead, which induces a voltage gradient $\nabla V$. Following the most conventional
 approach at low temperature regime, the electrical current density ${\bf \mathcal{J}} $ and
 the thermal current density ${\bf \mathcal{J}}_{q}$ can be written by following Onsager 
 relation\cite{onsagar1,onsagar2} as
\begin{equation}
{\bf \mathcal{J}} = Q^{11} {\bf \mathcal{E}} + Q^{12} (-{\nabla} T)
\end{equation}
and
\begin{equation}
{\bf \mathcal{J}}^{q} = Q^{21} {\bf \mathcal{E}} + Q^{22}(-{\nabla} T),
\end{equation}
where ${\bf \mathcal{E}} $ is the electric field and $ Q^{ij} $ ($ i,j=1,2$) are the 
phenomenological transport coefficients which can be expressed in terms of an integral 
$\mathcal{L}_{(\nu)}$: $Q^{11}=\mathcal{L}_{0}, Q^{21} = TQ^{12} = -\mathcal{L}_{1}/e$, 
$Q^{22}=\mathcal{L}_{2}/(e^2T)$ with
\begin{equation}
\mathcal{L}_{\nu} = \int dE  \Big[-\frac{\partial f(E)}{\partial E}\Big]
(E-\mu)^{\nu}  \mathcal{T}(E),
\end{equation}
where $\nu=0,1,2$ and $f(E)=[1+\exp(E-\mu)/k_BT]^{-1}$ is the Fermi-Dirac distribution
function with $\mu$ being the chemical potential. Here, $\mathcal{T}(E)$ is the energy-dependent
transmission amplitude. 
\begin{figure*}[!thpb]
\centering
\includegraphics[height=6cm,width=0.49\linewidth]{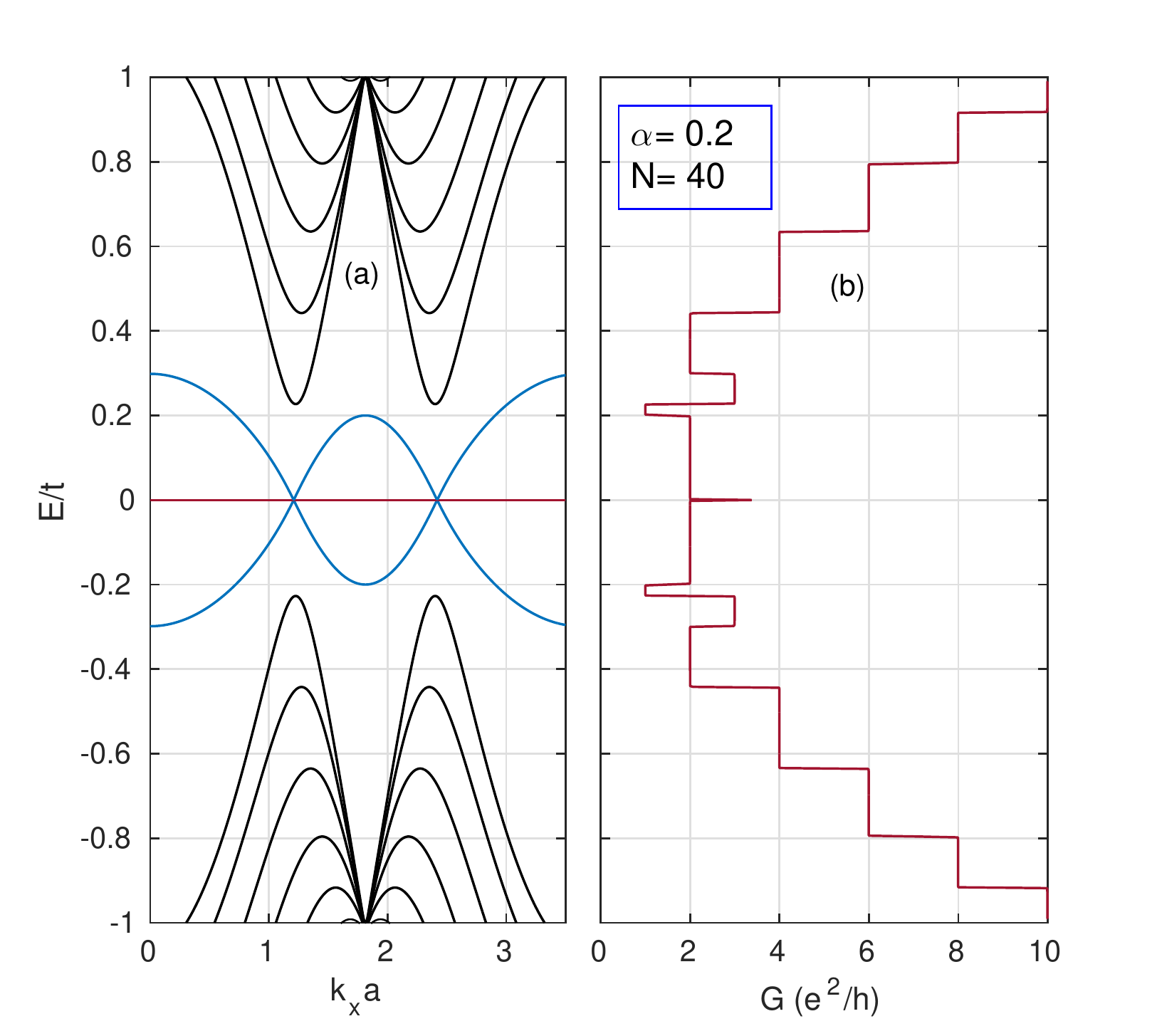}
\includegraphics[height=6cm,width=0.49\linewidth]{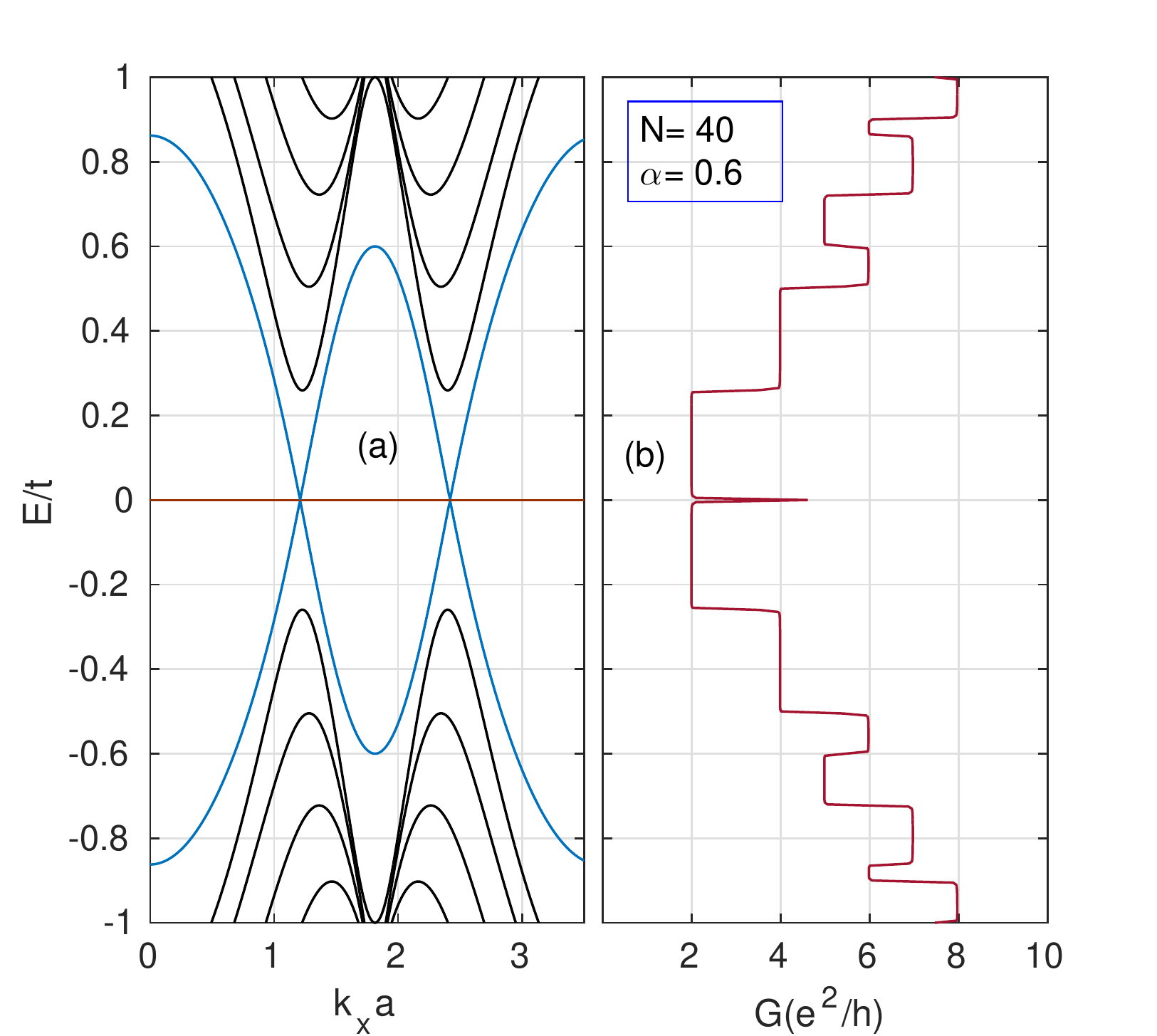}
\caption{conductance for (a) $\alpha=0.2$ and (b) $\alpha=0.6$. The width is $N=41$ in both cases.}
\label{con40_alpha}
\end{figure*}
Thermopower can be defined under open circuit condition
($\mathcal{J}=0$) as \cite{dutta,lv,groth,para_th,Ferrer,hatef}
\begin{equation}\label{power}
S=\frac{Q^{12}}{Q^{11}}=-\frac{1}{eT}\frac{\mathcal{L}_1}{\mathcal{L}_0}.
\end{equation}
On the other hand, the electronic contribution to the thermal conductance\cite{Ferrer,hatef} can be written as
\begin{equation}\label{TC}
\kappa_e=\frac{1}{hT}[\mathcal{L}_{2}-\mathcal{L}_1(\mathcal{L}_{0})^{-1}\mathcal{L}_{1}]
\end{equation}
The thermoelectric performance of a material is quantified by a parameter known as thermoelectric figure of merits
and it is given by\cite{Ferrer,hatef}
\begin{equation}
 ZT=\frac{S^2G_{T}T}{\kappa_e}=\frac{\mathcal{L}_1^2}{\mathcal{L}_0\mathcal{L}_2-\mathcal{L}_1^2}.
\end{equation}
Here, $G_{T}=(e^2/h)\mathcal{L}_0$ is the energy dependent electrical conductance following Landauer-Buttiker 
formula at non-zero temperature. Here, in the expression of ZT, thermal conductance is taken to be
electronic contribution only. The phonon/lattice contribution can be suppressed under very low or
very high temperature. One of the key ingredients in all the above equations is
the energy dependent transmission probability $\mathcal{T}(E)$. In order to obtain $\mathcal{T}(E)$ for a nanoribbon of this lattice, we shall use the
well known tight-binding Green function approach. We first give a brief review of this formalism
which was developed by Sancho\cite{sancho} to study the transfer matrices and spectral density
of states at the surface of a semi infinite crystal made of stacked layers. This approach can be
used in the hexagonal lattice too, where each supercell acts as independent layer. The method
has been already used in several hexagonal lattices like graphene\cite{para}, silicene\cite{silicene}, Mo$S_2$\cite{mos2}
, phosphorene\cite{njp_ezawa} etc. It is also worthwhile to mention at this stage that electron-electron
Coulomb interaction can play significant role in nanoribbon geometry which requires many-body Green function
GW approximation approach\cite{coulomb} or DFT method\cite{DFT} and will not be considered here. 

Our device is composed of three regions, the central region,
the left lead and right lead as shown in Fig.~(\ref{ribbon}). As the left and right leads
are identical, we can write $\mathcal{H}_{MM}=\mathcal{H}_{00}$ and $\mathcal{H}_{MM+1}=\mathcal{H}_{-10}$.
By implementing transfer matrix approach, the surface Green function can be written as
\begin{equation}
 \mathcal{G}_{00}^{L}(E)=[(E+i\eta)I-\mathcal{H}_{00}-\mathcal{H}^{\dagger}_{-10}\tilde{\Lambda}]^{-1},
\end{equation}
and
\begin{equation}
 \mathcal{G}_{00}^{R}(E)=[(E+i\eta)I-\mathcal{H}_{00}-\mathcal{H}^{\dagger}_{-10}\Lambda]^{-1},
\end{equation}
where $I$ is identity matrix. The notations $\Lambda$ and $\tilde{\Lambda}$ in above two equations can be evaluated as
 \begin{equation}\label{T}
 \Lambda=c_0+\tilde{c}_0c_1+\tilde{c}_0\tilde{c}_1c_2+.....+\tilde{c}_0\tilde{c}_1\tilde{c}_2...c_l
 \end{equation}
 and
 \begin{equation}\label{T1}
 \tilde{\Lambda}=\tilde{c}_0+c_0\tilde{c}_1+c_0c_1\tilde{c}_2+.....+c_0c_1c_2....\tilde{c}_l,
 \end{equation}
where $c_0$ and $\tilde{c}_0$ are defined as
\begin{equation}
 c_0=[(E+i\eta){\rm I}-\mathcal{H}_{00}]^{-1}\mathcal{H}^{\dagger}_{-10},
\end{equation}
and
\begin{equation}
 \tilde{c}_0=[(E+i\eta){\rm I}-\mathcal{H}_{00}]^{-1}\mathcal{H}_{-10}
\end{equation}
with
\begin{equation}
 c_i=({\rm I}-\tilde{c}_{i-1}c_{i-1}-c_{i-1}\tilde{c}_{i-1})^{-1}c_{i-1}^{2}
\end{equation}
and
\begin{equation}
 \tilde{c}_i=({\rm I}-\tilde{c}_{i-1}c_{i-1}-c_{i-1}\tilde{c}_{i-1})^{-1}\tilde{c}_{i-1}^{2}.
\end{equation}
\begin{figure*}[!thpb]
\centering
\includegraphics[height=6cm,width=0.49\linewidth]{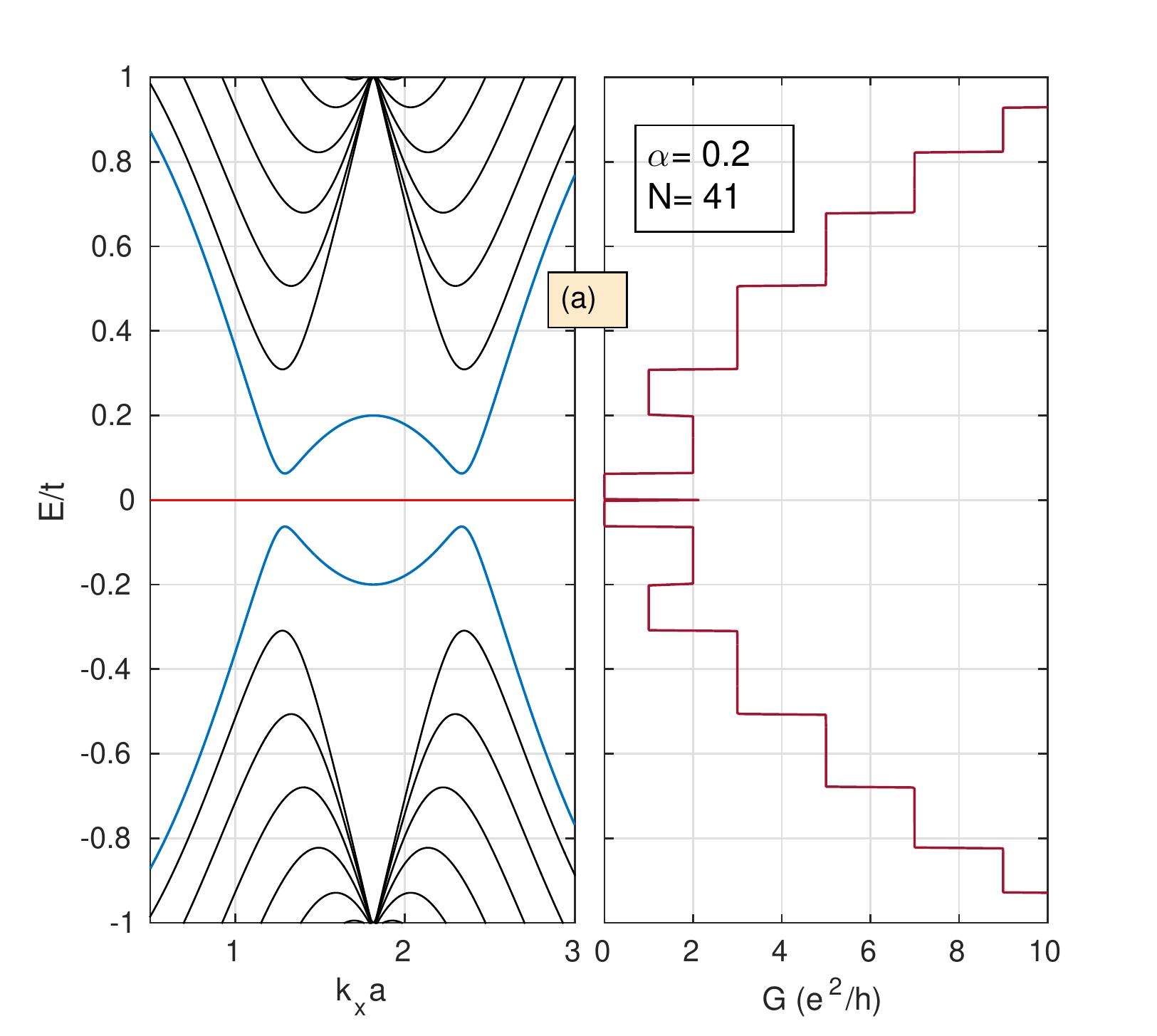}
\includegraphics[height=6cm,width=0.49\linewidth]{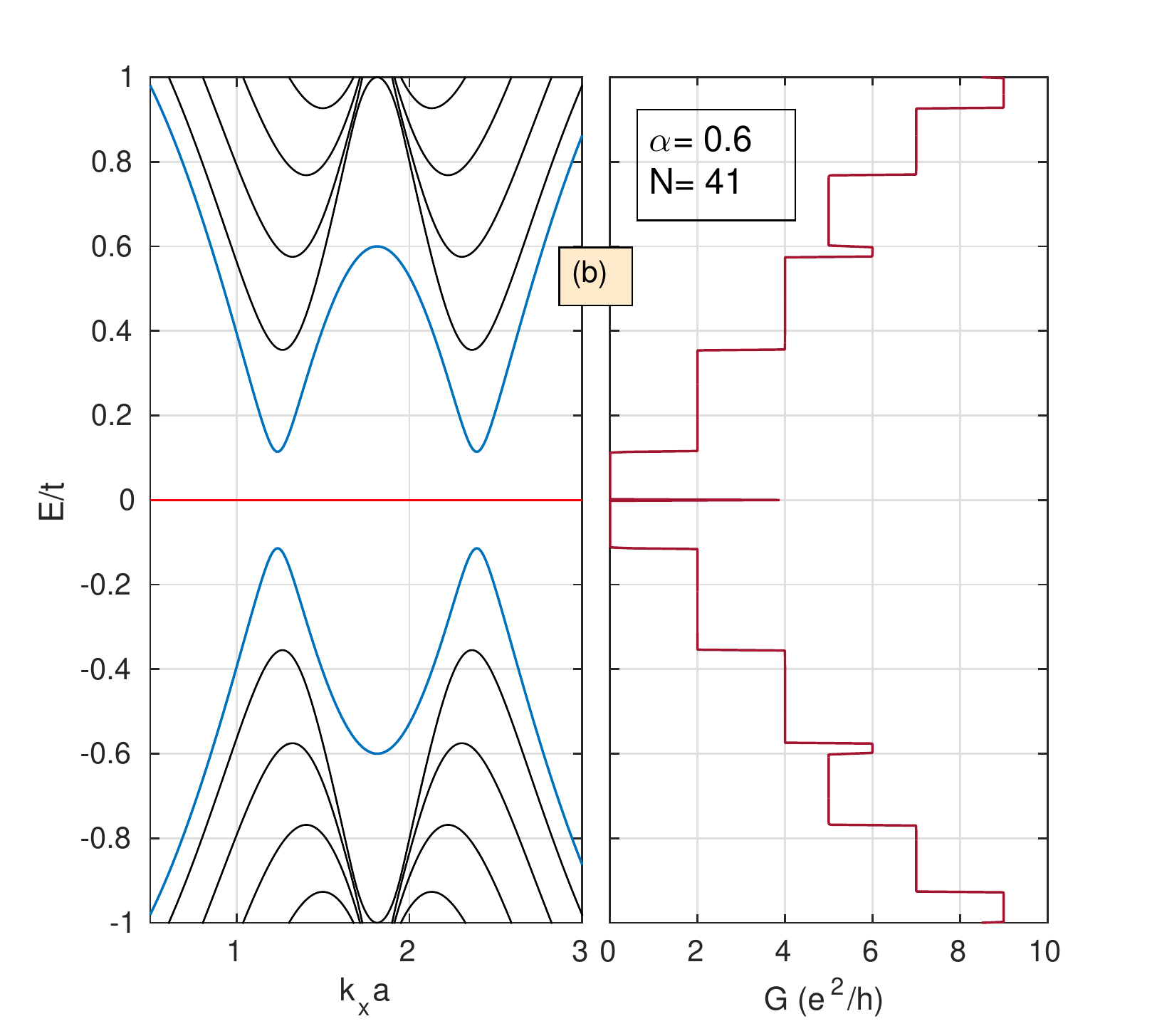}
\caption{conductance for (a) $\alpha=0.2$ and (b) $\alpha=0.6$. The width is $N=41$ in both cases.}
\label{con41_alpha}
\end{figure*}
The summation in Eqs.~(\ref{T}) and (\ref{T1}) has to be taken until $c_l$ and $\tilde{c}_l$ reach to zero.
The main advantage of this technique is that $2^l$ unit cells can be captured by just performing $l$ iteration.
Now we calculate surface Green function $\mathcal{G}_{22}$ by using the following recursion formula
\begin{equation}
 \mathcal{G}_{mm}^{R}=[(E+i\eta)I-\mathcal{H}_{mm}-\mathcal{H}_{mm+1}\mathcal{G}_{m+1m+1}^{R}\mathcal{H}^{\dagger}_{mm+1}]^{-1}.
\end{equation}
The effects of the left and right leads can be finally incorporated into the total Green function
via self energy as
\begin{equation}
 \mathcal{G}_{11}=[(E+i\eta)I-\mathcal{H}_{11}-\Sigma_{L}-\Sigma_R]^{-1}
\end{equation}
with
\begin{equation}
 \Sigma_{L}=\mathcal{H}_{01}^{\dagger}\mathcal{G}_{00}^{L}\mathcal{H}_{01}
\end{equation}
and
\begin{equation}
 \Sigma_{R}=\mathcal{H}_{12}\mathcal{G}_{22}^{R}\mathcal{H}_{12}^{\dagger}.
 \end{equation}
 Now we can define broadening matrix as
 \begin{equation}
  \Gamma_{L(R)}=i(\Sigma_{L(R)}-\Sigma^{\dagger}_{L(R)})
 \end{equation}
 which gives the transmission probability 
 \begin{equation}
  \mathcal{T}(E)=Tr[\Gamma_L\mathcal{G}_{11}\Gamma_R(\mathcal{G}_{11})^{\dagger}].
 \end{equation}
  Finally, using the Landauer-Buttiker formula, we obtain the electrical conductance as
 \begin{equation}\label{conduc}
  G=\frac{e^2}{h}\mathcal{T}(E)
 \end{equation}
 at zero temperature which subsequently leads to the
 temperature dependent conductance in terms of $\mathcal{L}_{\nu}$ as follows:
 \begin{equation}
  G_T=\frac{e^2}{h}\mathcal{L}_0,
 \end{equation}
\section{Numerical results and discussion}\label{sec4}
In this section, we present all transport coefficients numerically. First, we evaluate the conductance
by using Eq.~(\ref{conduc}) and plot it with respect to the energy dispersion for
$\alpha=0.2$ in Fig.~(\ref{con40_alpha})a. Here, we keep the width $N=3q+1$ with $q$
being the positive integer, in order to capture the gapless chiral edge modes.
The conductance appears to be quantized as $2re^2/h$ (r being the positive integer) with
the `$2$' factor attributes to the two valleys. The integer `r' accounts the number of
transverse modes (black line) including the edge modes. With the increase of chemical potential,
transverse modes start to penetrate through the chemical potential one by one, leading to the
increase of conductance. Each transverse modes contributes conductance by $2$ units.
\begin{figure}[htb]
\begin{minipage}[t]{0.5\textwidth}
 {\includegraphics[width=.5\textwidth,height=5cm]{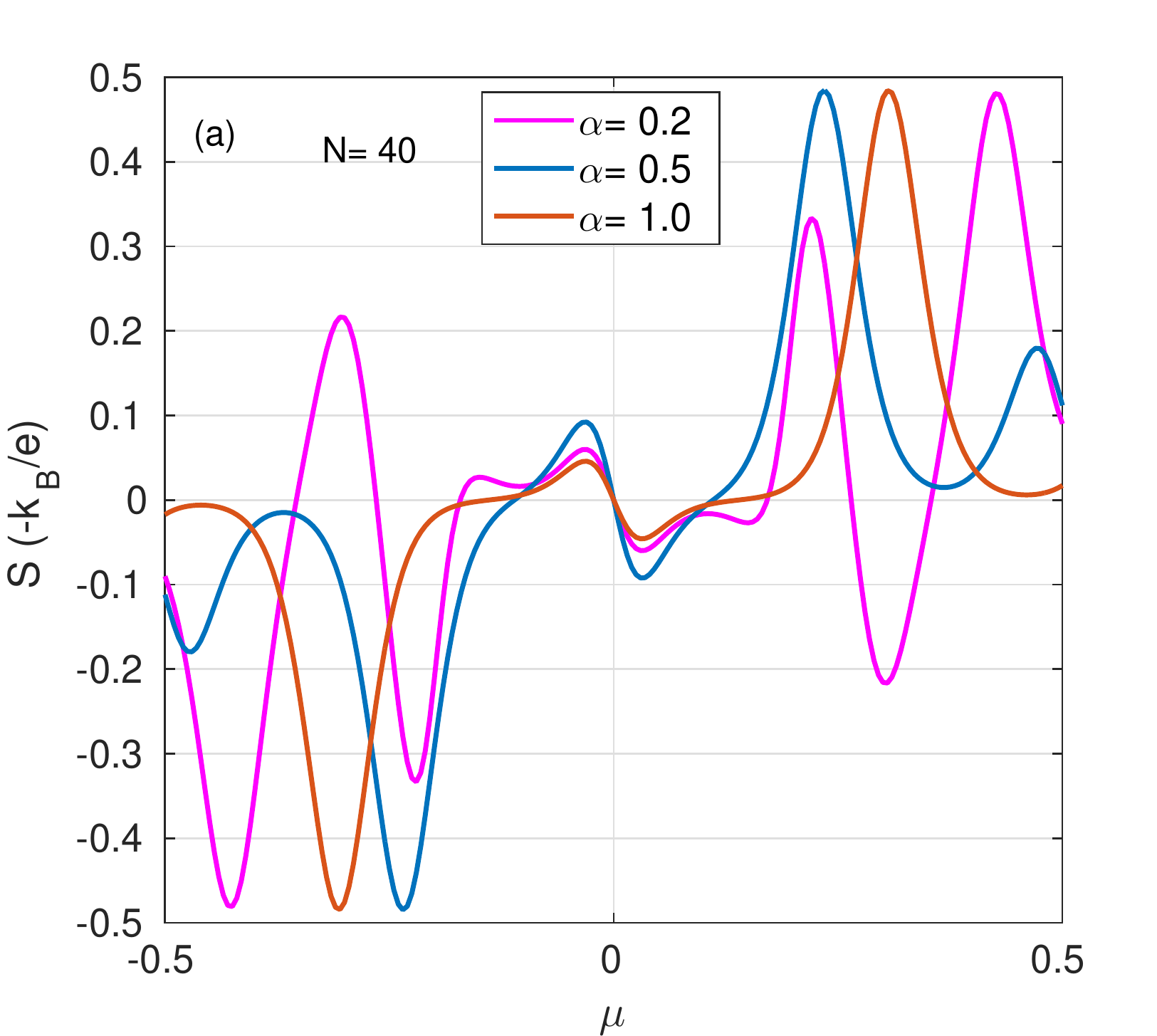}}
 \hspace{-.2cm}{ \includegraphics[width=.5\textwidth,height=5cm]{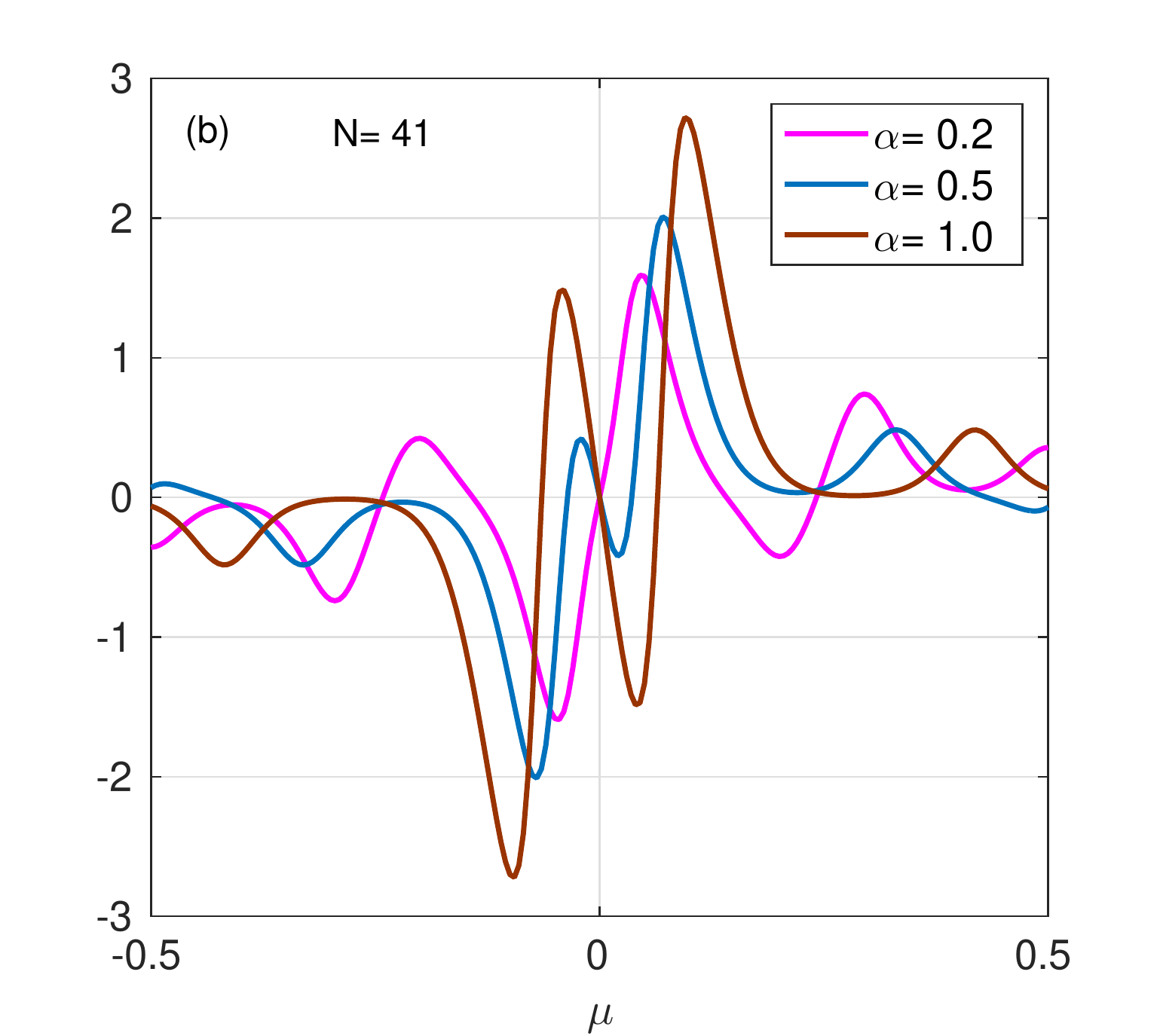}}
 \end{minipage}
 \caption{Thermopower as a function of chemical potential for (a) $N=40$ and (b) $N=41$.
 The temperature is taken at $T=0.02\frac{t}{k_B}$.}
\label{thermopower}
\end{figure}
However, an unusual behavior is also seen in the region $-0.4\le E/t\le 0.4$ where edge modes (sky blue line)
reside. Note that the pair of edge modes gives rise to the quantized conductance of $2e^2/h$ unit before stepping
down by one unit. This happen at a region where edge modes and 1st transverse modes interfere each other.
A central peak in the conductance emerges corresponding to the dispersionless flat band.
Similar peak resembles the divergence in dc bulk conductivity of such lattice under clean limit due to interband
scattering\cite{mate}, at the band touching point between the flat band and the conic band. We find
almost similar feature for $\alpha=0.6$, as shown in Fig.~(\ref{con41_alpha})b, except edge modes
contribute for a wide range of energy.
\begin{figure}[!thpb]
\centering
\includegraphics[height=7cm,width=\linewidth]{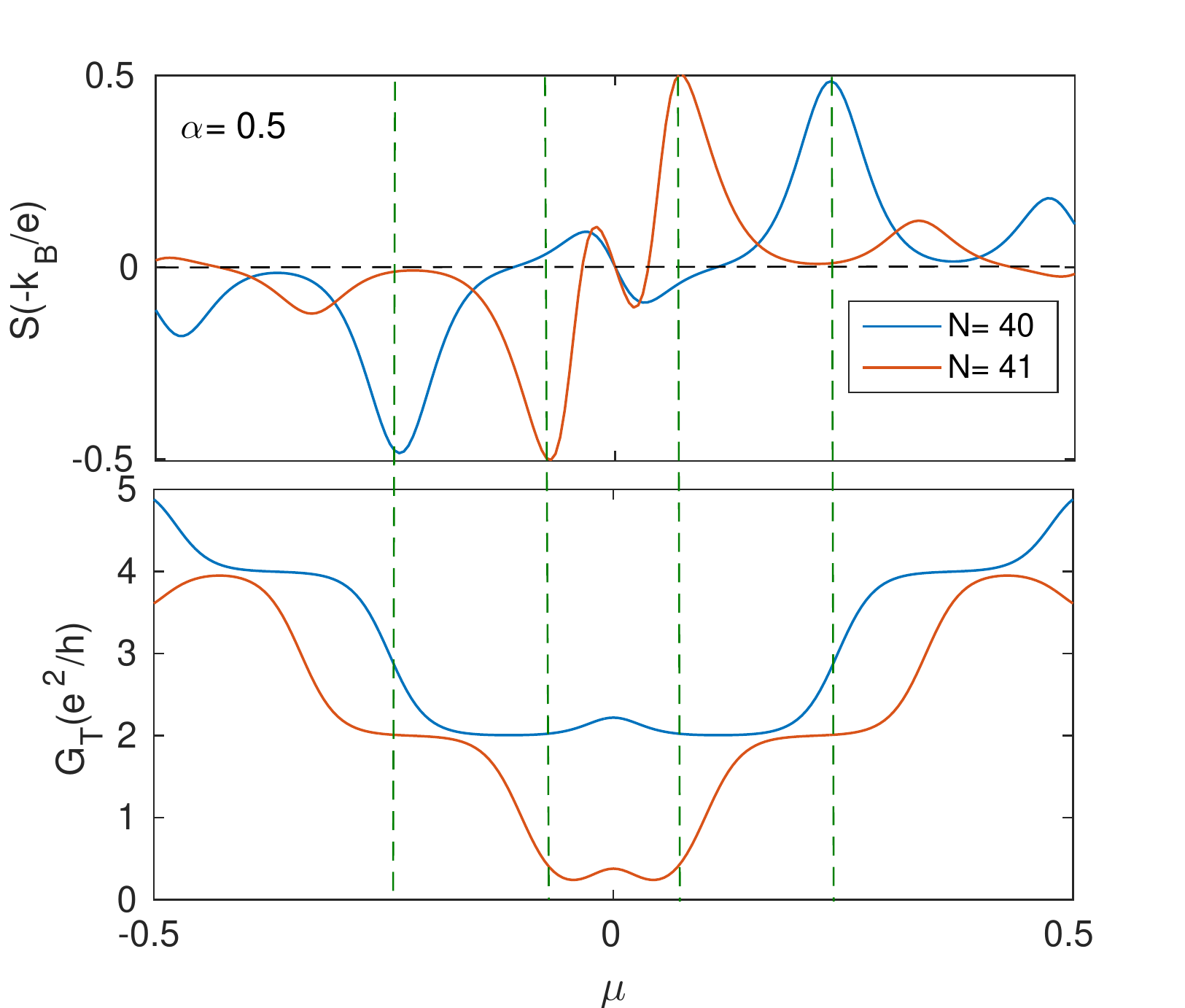}
\caption{Thermopower (upper panel) and conductance(lower panel) versus chemical potential. Thermopower for $N=41$
is scaled down by a factor $1/4$.}
\label{SpG}
\end{figure}

Now we explore how the conductance gets affected by changing $\alpha$ for the gapped dispersion.
To keep the band dispersion gapped, we take the width of the ribbon $N\ne3q+1$ ($41$ and $43$).
The conductance is plotted against the gapped energy dispersion as a function of chemical
potential in Fig.~(\ref{con41_alpha}). It is already shown in Fig.~(\ref{alpha_forN41}) that
the band gap increases slowly with the increase of $\alpha$. This fact has a direct impact on
the conductance, as shown in Fig.~(\ref{con41_alpha}), in terms of widening the zero conductance
region with the increase of $\alpha$. This is a direct signature of the gap opening in transport measurement in
a zigzag edge nanoribbon of such material provided the width has to be other than $N=3q+1$.
Note that the degenerate flat bands also induce a central peak in the conductance spectra,
however its height varies with the strength of $\alpha$. Another noticeable point here is
that although the conductance steps down by unity in case of $\alpha=0.2$ at $E/t=\pm 0.2$,
it disappears for $\alpha=0.6$. The origin of it can be attributed to the peculiar feature 
of the edge modes in the region $1.2<k_xa<2.5$ in both cases. 
\begin{figure}[htb]
\begin{minipage}[t]{0.5\textwidth}
 {\includegraphics[width=.5\textwidth,height=5cm]{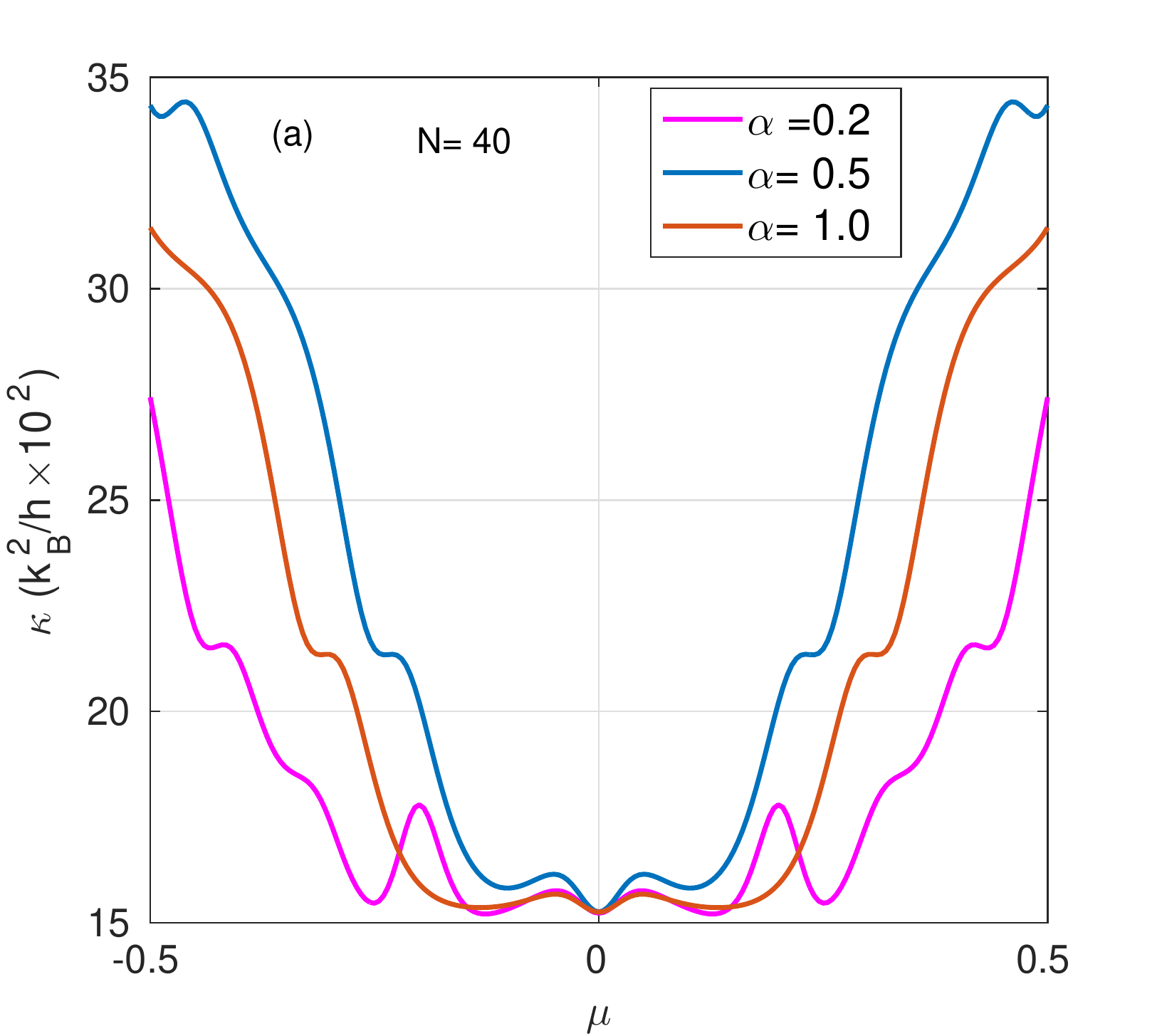}}
 \hspace{-.2cm}{\includegraphics[width=.5\textwidth,height=5cm]{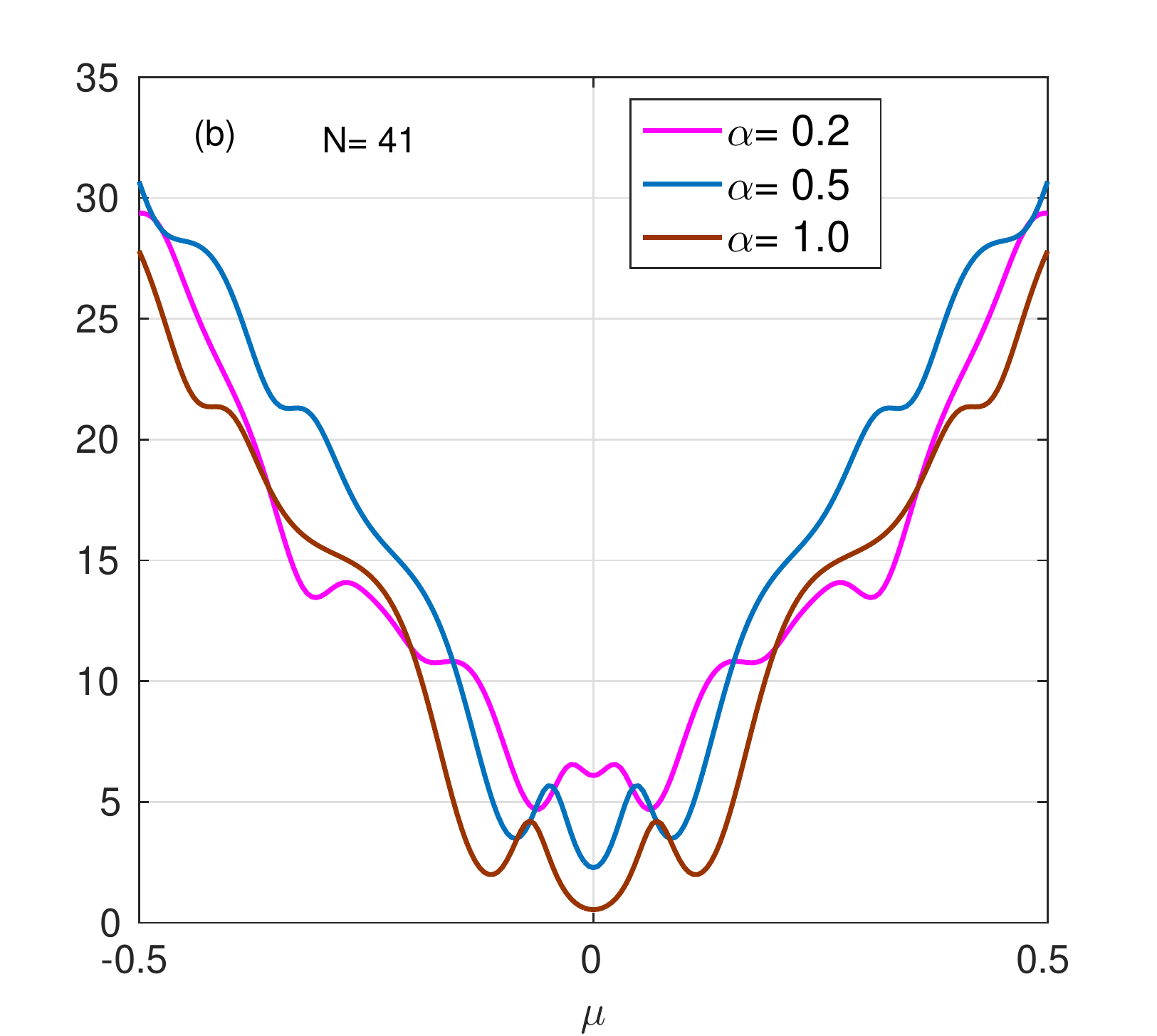}}
 \end{minipage}
 \caption{Thermal conductance versus chemical potential for (a) $N=40$ and $N=41$.
 The temperature is taken at $T=0.02\frac{t}{k_B}$.}
\label{th_con}
\end{figure}

Now we plot thermopower by using Eq.~(\ref{power}) in Fig.~(\ref{thermopower}). We consider 
both cases i.e., gapless and gapped dispersion by choosing the width $N=40$ and $N=41$.

The thermopower for the width $N=40$ is plotted in Fig.~(\ref{thermopower})a which shows that non-zero $\alpha$
has no much significant impact except the shifting of the thermopower peaks. However, a significant
enhancement of thermopower can be found for gapped dispersion as can be seen in Fig.~(\ref{thermopower})b.
It is worthwhile to mention that the thermopower can be further linked to the conductance
via the standard Mott's relation 
\begin{equation}
S\sim-\left[\frac{d}{dE}ln[G_T(E)]\right]_{E=\mu}
=-\left[\frac{1}{G_{T}(E)}\frac{d}{dE}G_{T}(E)\right]_{E=\mu}
\end{equation}
which indicates that thermopower should be maximum around the slope of conductance spectrum,
which can be confirmed from the plot Fig.~(\ref{SpG}). At the same time,
the amplitude of thermopower decreases with $G_{T}(E)$ as they are inversely related.
On the other hand, it is noted from Fig.~(\ref{thermopower})b that thermopower increases with the strength
of $\alpha$. Note that for gapped edge modes ($N=41$), the thermopower is enhanced with the increase of $\alpha$,
where as such enhancement does not occur with $\alpha$ for gapless edge modes ($N=40$). 
The reason can be attributed to how $\alpha$ affects the product of the slope
of conductance spectrum and it's inverse. The enhancement of thermopower with the $\alpha$ 
reveals that this product is much sensitive to the $\alpha$ for gapped edge modes
in comparison to the gapless edge modes.
\begin{figure}[htb]
\begin{minipage}[t]{0.5\textwidth}
{\includegraphics[width=.45\textwidth,height=5cm]{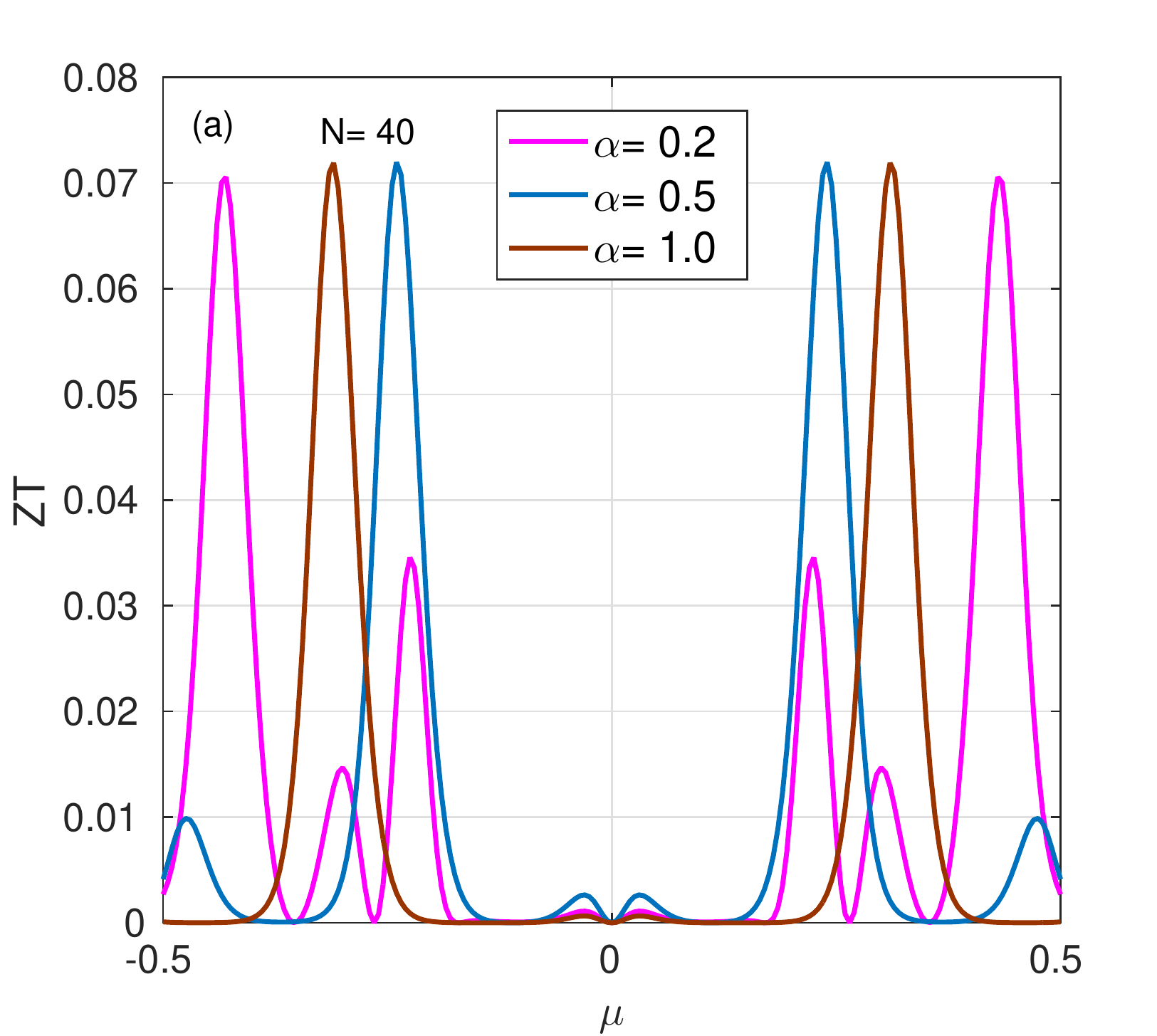}}
\hspace{-.2cm}{\includegraphics[width=.45\textwidth,height=5cm]{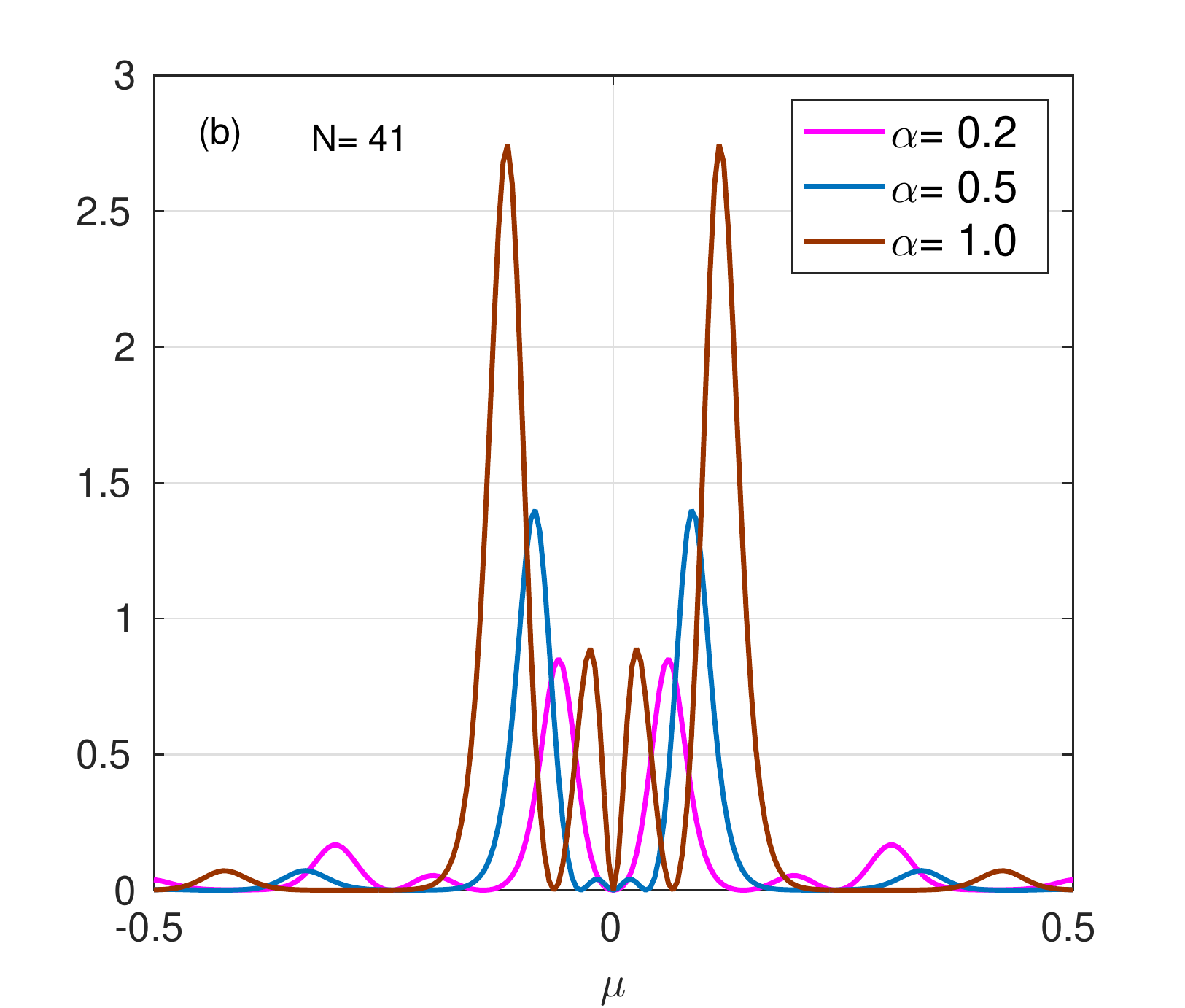}}
\end{minipage}
\caption{Thermoelectric figure of merits (ZT) versus the chemical potential (a) $N=40$ and (b) $N=41$ 
at $T=0.02t\frac{e}{k_B}$.}
\label{merits}
\end{figure}
The corresponding thermal conductance are also evaluated by using Eq.~(\ref{TC}) and plotted
in Fig.~(\ref{th_con}). The Fig.~(\ref{th_con}) shows that unlike the case of thermopower,
the thermal conductance is relatively much less sensitive to the variation of $\alpha$. The Fig.~(\ref{TC})a
is plotted for the gapless edge modes i.e., $N=40$. It shows that the effect of $\alpha$ is
relatively stronger for doped ribbon. On the other hand, for gapped system, effects of $\alpha$
seems to be stronger around undoped situation. 

Finally, we look whether the dice lattice (or $T_3$) can of a way to improve the 
thermoelectric figure of merits of such hexagonal lattice. In order to address this concern, the
thermoelectric figure of merit is plotted as a function of $\alpha$ for two different widths in Fig.~(\ref{merits}).
It shows that the maximum value of the thermoelectric figure of merits remains unaltered with the 
variation of $\alpha$ except a shift of the peaks for gapless edge modes (width $N=40$). However, on the 
other hand the figure of merits for $N=41$ is plotted in Fig.~(\ref{merits})a (gapped dispersion)
which shows that figure of merits gets almost thirty times larger than than $N=40$ (\ref{merits})b
(gapless dispersion). We can conclude that the zigzag ribbon of a $T_3$/dice lattice with 
gapped dispersion can be a better choice for thermoelectric material.

{\it{\bf Line defects and its consequences:}}\\
The line defects in honeycomb lattice are formed or created out of the absence of one or more sublattices
in each unit cell\cite{defects1,defects2}. The presence of vacancy/line defects is one of the most common 
issue which arises during experimental realization of such lattice. It has been previously 
observed that such line defects yield some important consequences on the band structure
as well as in transport phenomena, such as opening the gap in graphene\cite{peeters}, valley polarization\cite{white} etc
 \begin{figure}[!thpb]
\begin{minipage}[t]{0.5\textwidth}
  \hspace{-.4cm}{ \includegraphics[width=.5\textwidth,height=4cm]{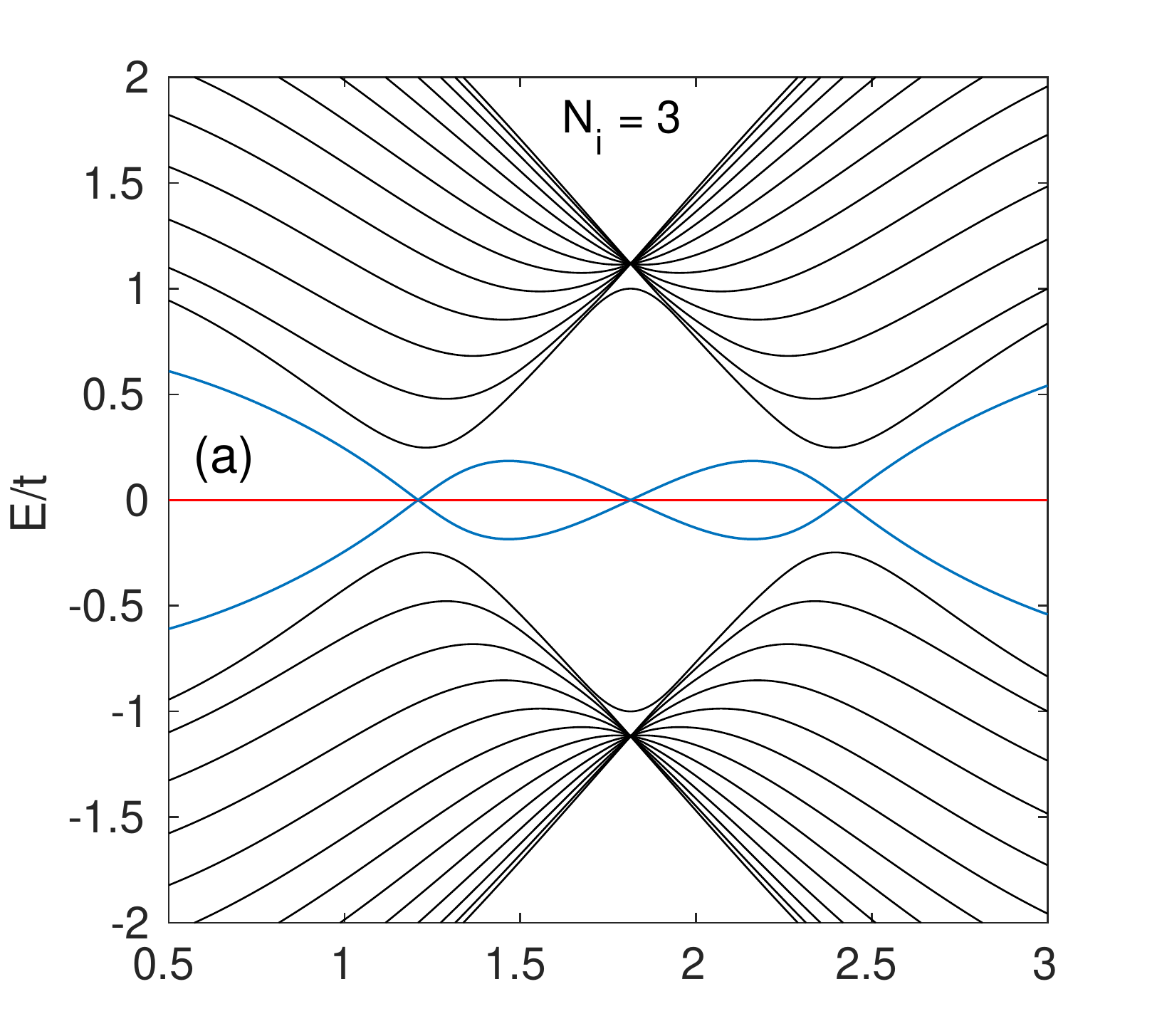}}
  \hspace{-.5cm}{ \includegraphics[width=.5\textwidth,height=4cm]{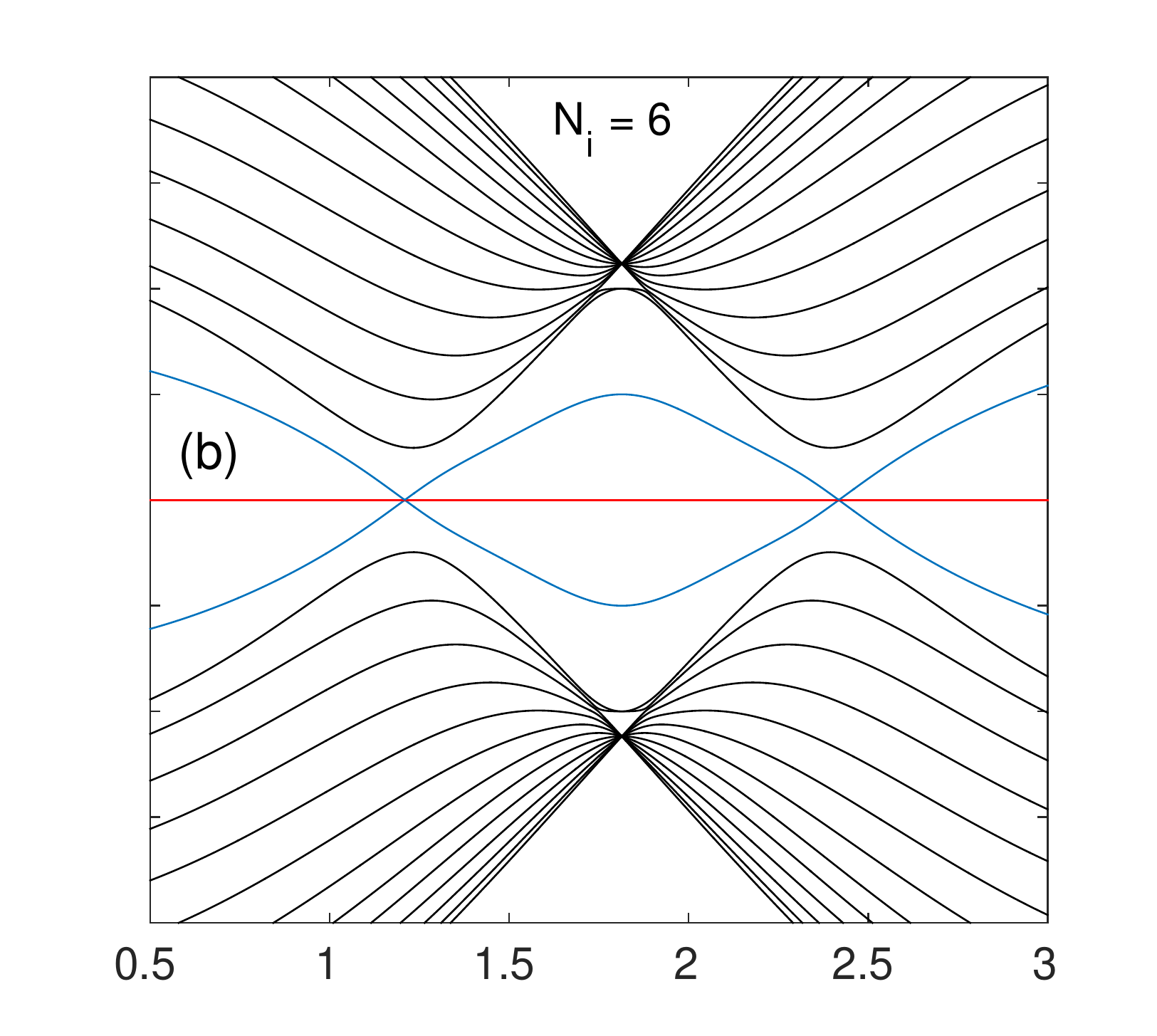}}
\end{minipage}
\begin{minipage}[t]{0.5\textwidth}
  \hspace{-.4cm}{ \includegraphics[width=.5\textwidth,height=4cm]{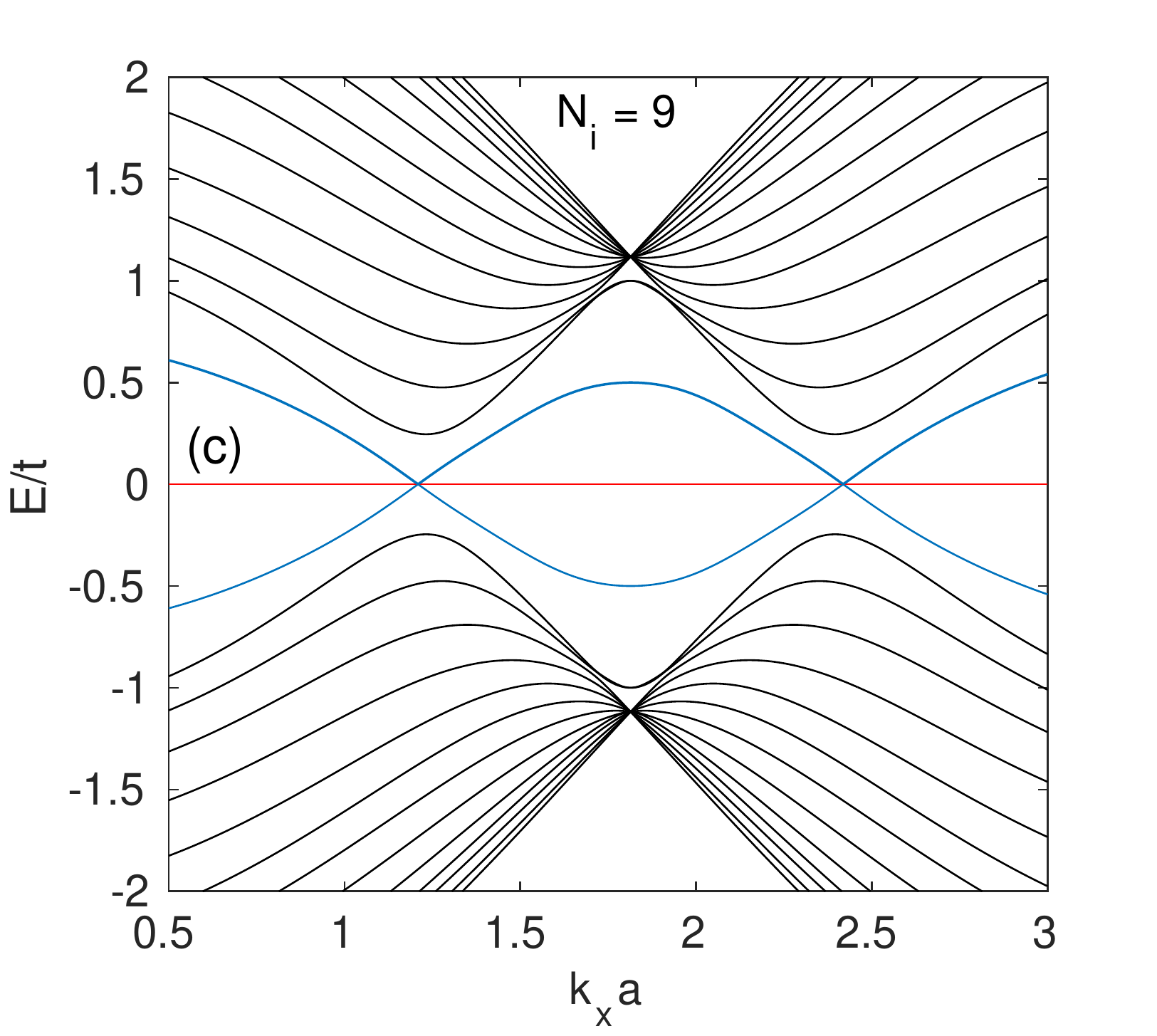}}
  \hspace{-.5cm}{ \includegraphics[width=.5\textwidth,height=4cm]{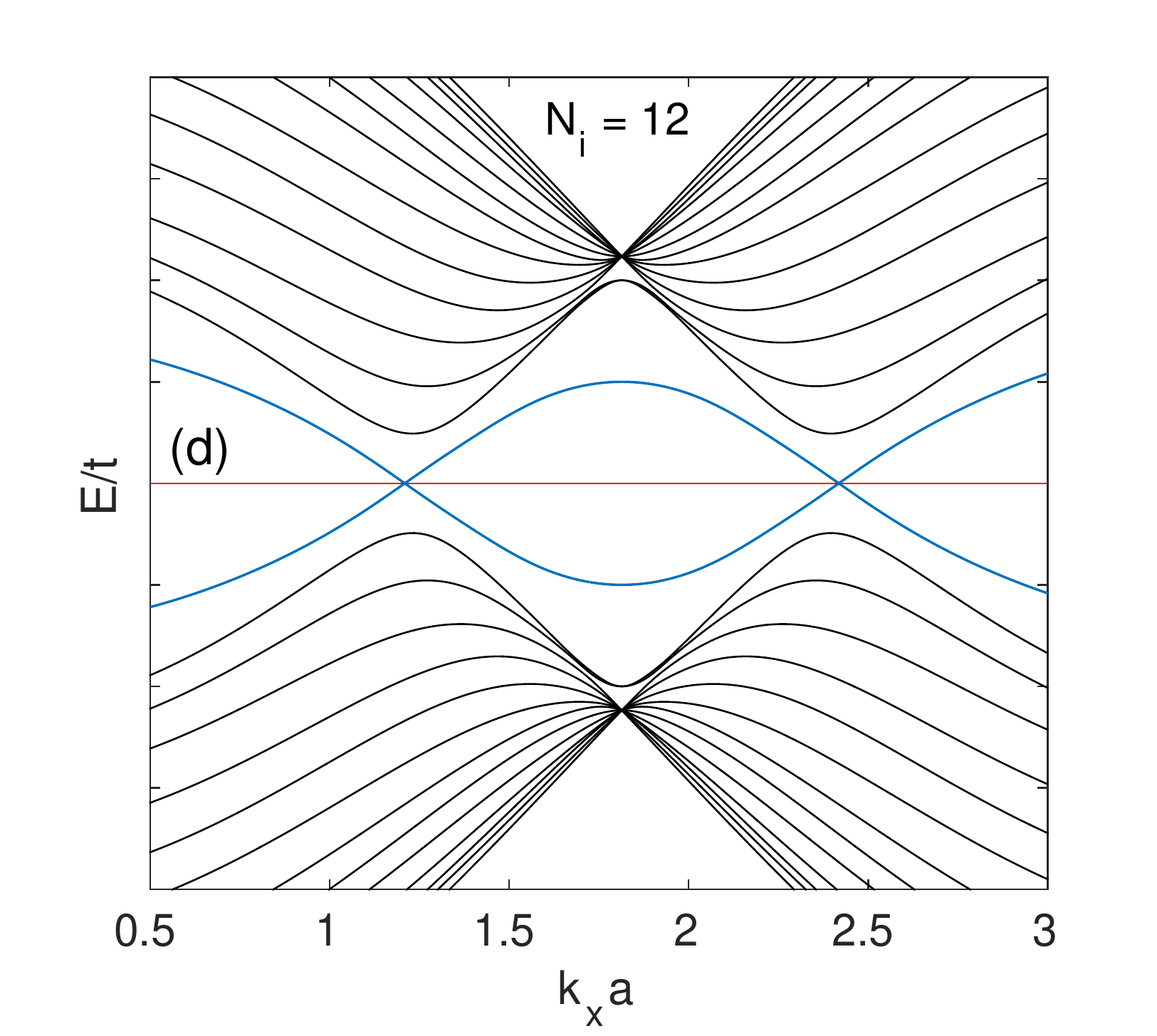}}
\end{minipage}
 \caption{Energy dispersion of the zigzag edge nanoribbon in presence if line defects.
 Four cases are considered here. Line defects are inserted by removing $C$ sublattices
 at the i-th position i.e., (a) $N_i=3$ (b) $N_i=6$ (c) $N_i=9$ and (d) $N_i= 12$. The width of the ribbon
 is $N=40$ and $\alpha= 0.4$ for all cases.}
 \label{band_defects}
 \end{figure}
 
In this section, we examine the effects of the line defects, formed out of the absence of $C/A$
sublattices in each unit cell at different distances from the edge. The effects of line
defects/vacancy out of the absence of $A$ or $B$ sublattices were investigated
previously in graphene\cite{para,peeters} or silicene\cite{silicene}. In order to incorporate the line defect
in tight-binding calculation, we set on-site energy to infinity at the missing site which prevents
hoping between vacancy to nearest sites. The band dispersion of zigzag nanoribbon in presence of line
defects are numerically presented in Fig.~(\ref{band_defects}). The Fig.~(\ref{band_defects})a
is plotted for line defects created out of the absence of $C$ atoms (C atoms) at $N_i=3$ where `$i$' is
the sublattice number index with $1\le i\le N$. It shows that the line defect causes drastic 
changes to the feature of the edge modes by enforcing them to touch at the middle of the two
Dirac points. This feature can be expected to have significance consequences on the transport 
properties.

On other hand if the line defect is situated away from the zigzag edge
i.e for distance $N_i=6,9,12$, the effects on band spectrum appears to be almost negligible
[see Fig.~(\ref{band_defects}))b-(\ref{band_defects})d].
\begin{figure}[htb]
\begin{minipage}[t]{0.5\textwidth}
{\includegraphics[width=.46\textwidth,height=4.5cm]{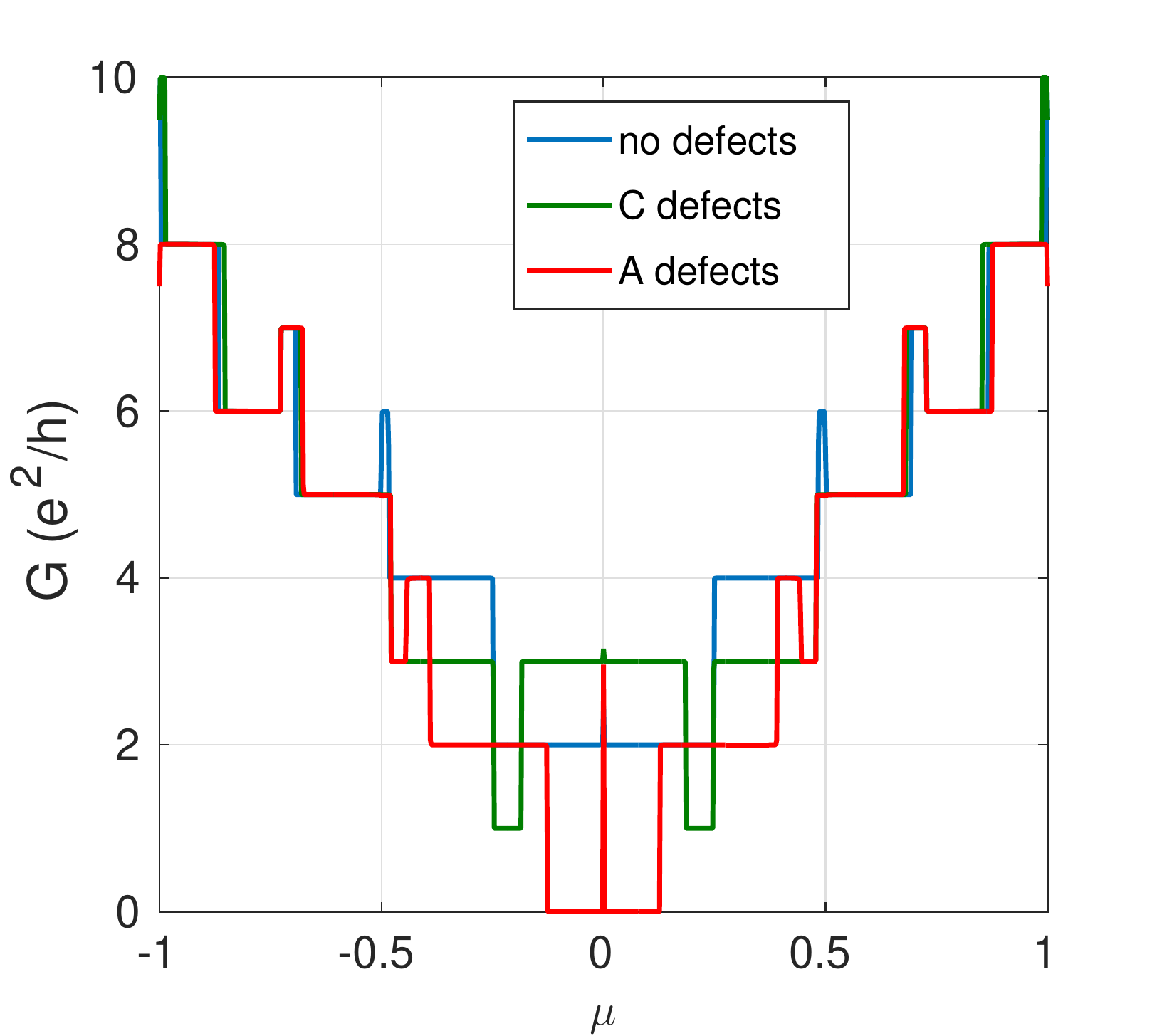}}
\hspace{-0.15cm}{\includegraphics[width=.46\textwidth,height=4.5cm]{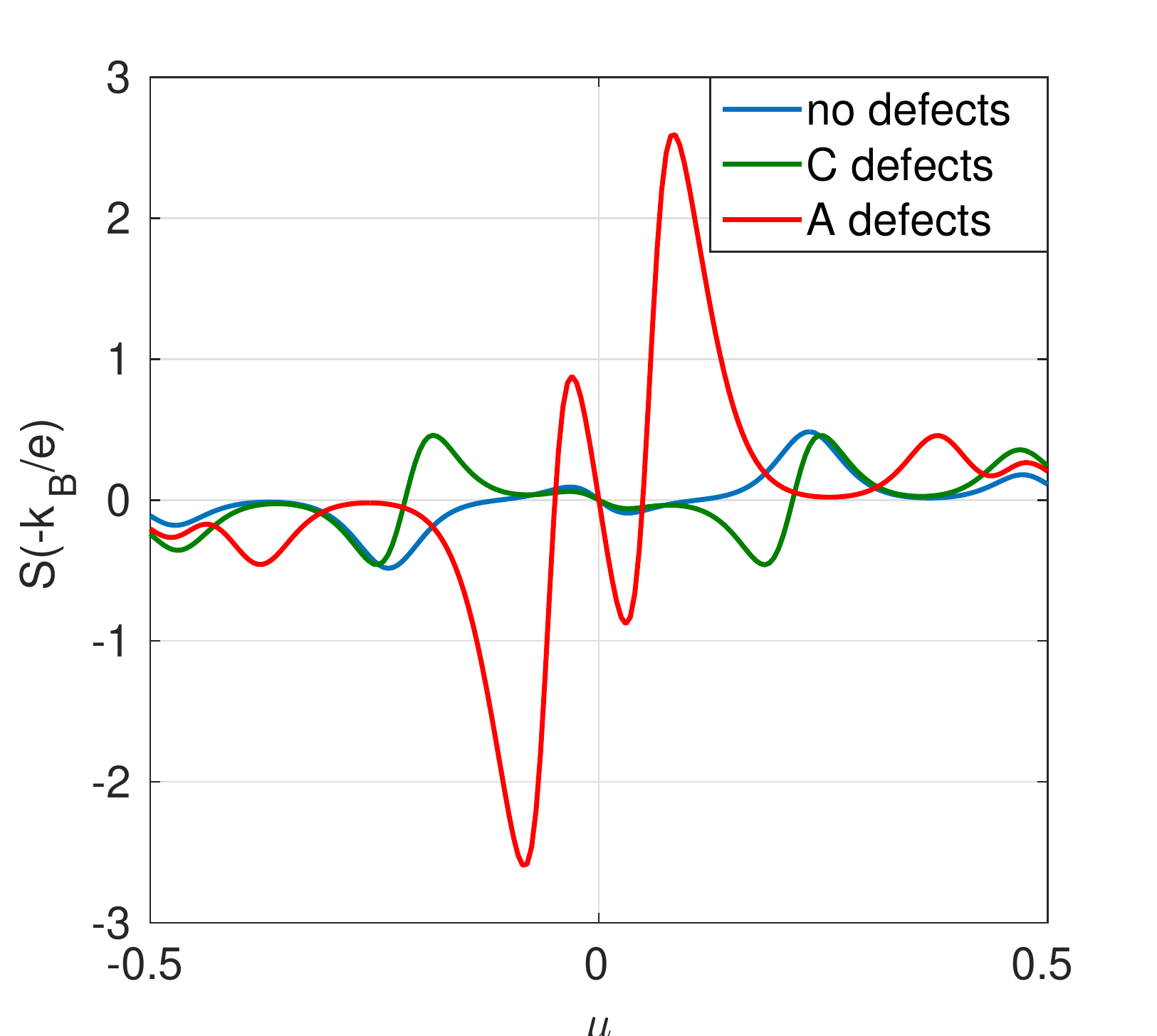}}
\end{minipage}
\caption{Conductance and thermopower are plotted versus chemical potential in (a) and (b), respectively.
The line defect is considered at $N_i=3$ for c defects and $N_i=4$ for A defects.
The width of the ribbon is $N= 40$ and $\alpha= 0.5$.}
\label{defect_con}
\end{figure}

In order to probe the consequences of the effects of line defects on transport properties, we 
also plot conductance in Fig.~(\ref{defect_con}) for the line defect situated nearest to the edge
i.e., $N_i=3$. The most important signature of such line defect in the conductance is the 
appearance of conductance by $3e^2/h$ instead of $2e^2/h$ without line defects at low energy regime.
The origin can be traced to the band dispersion in Fig.~(\ref{band_defects})a which exhibits extra Dirac-like
point in addition to the two Dirac points, resulting three units of conductance. On the other hand,
in absence of line defects, it is the two valleys which contribute two units of conductance in the 
low energy regime. A peculiar behavior occurs for $\mu>0.2$ where the conductance steps down and
steps up by $2e^2/h$ without defects. Such contradictory features in both cases can be attributed
to the ways edge modes interfere with the transverse modes. We have also included the effects of 
line defects, formed out of the absence of A sublattices (A defects) close to the edge, on the 
conductance (shown in red line). It also exhibits quite distinct features in comparison to the 
C defects, in terms of the appearance of zero conductance on both side of the zero chemical potential.
It suggests that A defects can induce a small gap in the band dispersion too.
Subsequently, we also plot the thermopower in Fig.~(\ref{defect_con})b to reveal the effects
of the nearest line defect. It is observed that although at low energy regime conductance
gets affected by the presence of line defect, the thermopower seems to be lees sensitive
to the C defects in terms of the amplitude. However, the A defects enhance the thermopower
significantly.

As it is already shown in Fig.~(\ref{band_defects})b to (\ref{band_defects})d that line
defects, situated away from the edge has very less consequences and hence it can easily
be anticipated that such defects would not have any significant impacts on electric or
thermoelectric transport properties.

Finally we quickly comment here that the random disorder can be also treated in similar fashion by
incorporating on-site potential, distributed randomly in the system. However, we have already noticed that
the line defects can affect the feature of edge modes and corresponding
transport signature only if it resides close to the edge.
So we can conclude that the presence of random disorders can not have much significant impact
unless it resides close to the edge.
\section{Summary and conclusions}\label{sec5}
In this work, we explore the roles of zigzag edge geometry of $\alpha-\mathcal{T}_3$ lattice 
on the band dispersion, conductance, thermopower and thermoelectric figure of merits under
the continuous evolution of from graphene to dice lattice (by means of tuning $\alpha$).
We notice that the feature of edge modes are very much sensitive to the width of the nanoribbon.
The energy dispersion can be gapped or gapless depending on the width of the ribbon.
The edge modes are not dispersionless flat as found in graphene, rather it can be gapless 
chiral at the two valleys for specific width $N=3q+1$. Additionally, the slope of the gapless
chiral edge modes increases with the increase of $\alpha$. On the other hand, the gap opening occurs
between the pair of edge modes for the width of $N\ne3q+1$ and the energy gap increases
with the evolution towards dice lattice. Subsequently, we use tight-binding Green function approach to
analyses the roles of edge modes and $\alpha$ on electrical and thermoelectric transport coefficients
of the zigzag nanoribbon based device, attached to the left and right lead. We found that 
possibility of reshaping the edge modes, by means of width and $\alpha$, can be exploited to
improve the thermoelectric performances of such materials. It is found that the thermopower
and thermoelectric figure of merits can be enhanced significantly by means of $\alpha$.
The thermal conductance remains less sensitive to the $\alpha$ in comparison to thermopower whereas
the figure of merits exhibits a sharp enhancement. 
Finally, we have studied the consequences of line defects out of the absence of $C$ sublattices. We have found 
that such line defect has too weak impact on the band structure as well as on transport properties as long as
the defects resides away from the edge. However, there is a drastic changes in the nature of the edge modes
and corresponding transport signature if the line defects reside very near to the edge.
\begin{acknowledgements}
The authors thank the Deanship of Scientific Research in King Faisal University (Saudi Arabia) for funding the 
facilities required for this research as part of the Research Grants Program Nasher: 186124)
\end{acknowledgements}


\begin{thebibliography}{99}

\bibitem{novo} K. S. Novoselov, A. K. Geim, S. Morozov, D. Jiang, Y. Zhang,
               S. Dubonos, I. Grigorieva, and A. A. Firsov, Science \textbf{306}, 666 (2004).
\bibitem{neto} A. H. Castro Neto, F. Guinea, N. M. R. Peres, K. S. Novoselov,
                and A. K. Geim, Rev. Mod. Phys. \textbf{81}, 109 (2009).  
\bibitem{vidal} J. Vidal, R. Mosseri, and B. Doucot, Phys. Rev. Lett. \textbf{81}, 5888 (1998).

\bibitem{malcolm} J. D. Malcolm and E. J. Nicol, Phys. Rev. B \textbf{92}, 035118 (2015).

\bibitem{janik} Balazs Dora, Janik Kailasvuori, and R. Moessner, Phys. Rev. B {\bf 84}, 195422 (2011).

\bibitem{lan} Z. Lan, N. Goldman, A. Bermudez, W. Lu, and P. Ohberg, Phys. Rev. B {\bf 84}, 165115 (2011).

\bibitem{morigi} A. Raoux, M. Morigi, J.-N. Fuchs, F. Pi ́echon, and G. Montambaux, Phys. Rev. Lett. {\bf 112}, 
                 026402 (2014). 

\bibitem{Dxiao} D. Xiao, M. Chang, and Q. Niu, Rev. Mod. Phys. \textbf{82}, 1959 (2010).

\bibitem{nicol1} E. Illes, J. P. Carbotte, and E. J. Nicol, Phys. Rev. B {\bf 92}, 245410 (2015).

\bibitem{tutul1} Tutul Biswas and T. K. Ghosh, J. Phys.: Condens. Matter {\bf 28}, 495302 (2016).

\bibitem{firoz_para} SK F. Islam and P. Dutta, Phys. Rev. B {\bf 96}, 045418 (2017).

\bibitem{daniel} F. U.  Daniel, D. Bercioux, M. Wimmer and W. Hausler, Phys. Rev. B \textbf{84}, 115136 (2011).

\bibitem{klein} E. Elles and E. J. Nicol, Phys. Rev. B \textbf{95}, 235432 (2017)

\bibitem{nicol3} E. Illes, and E. J. Nicol, Phys. Rev. B \textbf{94}, 125435 (2016).

\bibitem{dora3} A. D. Kovacs, G. D. B. Dora, and J. Cserti, Phys. Rev. B \textbf{95}, 035414 (2017).

\bibitem{plasmon} J. D. Malcolm and E. J. Nicol, Phys. Rev. B {\bf 93}, 165433 (2016).

\bibitem{newT3} Y. R. Chen et al., Phys. Rev. B \textbf{99}, 045420 (2019).

\bibitem{photo} B. Dey and T. K. Ghosh, Phys. Rev. B \textbf{98}, 075422 (2018).

\bibitem{dora2} B. Dora, I. F. Herbut, and R. Moessner, Phys. Rev. B \textbf{90}, 045310 (2014)

\bibitem{tutul2} T. Biswas and T. K. Ghosh, J. phys.: Condens. Matter \textbf{30}, 075301 (2018).

\bibitem{nolas} Nolas G S, Sharp J and Goldsmid H J, 2001 Thermoelectrics (Berlin: Springer)

\bibitem{disalvo} F. J. DiSalvo, Science 285 703 (1999).

\bibitem{snyder} G. J. Snyder and E. S. Toberer, Nature Mater. 7 105 (2008).

\bibitem{das_sarma} E. H. Hwang, E. Rossi and S. D. Sarma, Phys. Rev. B \textbf{80}, 235415 (2009).

\bibitem{nam} S.-G. Nam, D. K. Ki, and H. J. Lee, Phys. Rev. B \textbf{82}, 245416 (2010).

\bibitem{hao} L. Hao and T. K. Lee, Phys. Rev. B \textbf{81}, 165445 (2010).

\bibitem{yuri} Yuri M. Zuev, W. Chang and P. Kim, Phys. Rev. Lett. \textbf{102}, 096807 (2009).

\bibitem{wei} P. Wei, W. Bao, Y. Pu, C. N. Lau, and J. Shi, Phys. Rev. Lett. \textbf{102} 166808 (2009).

\bibitem{hossain} M. S. Hossain, F. A. Dirini, F. Hossain and E. Skifidas, Sci. Rep. \textbf{5} 11297 (2015).

\bibitem{van} V.-T. Tran et al., Sci. Rep. \textbf{7}, 2313 (2017)

\bibitem{Ma} R. Ma, H. Geng, W. Y. Deng, M. N. Chen, L. Sheng, and D. Y. Xing,
Phys. Rev. B \textbf{94}, 125410 (2016).

\bibitem{flores} E. Flores, J. R. Ares, A. Castellanos-Gomez, M. Barawi, I. J. Ferrer, and
                 C. Sánchez, Appl. Phys. Lett. \textbf{106}, 022102 (2015).
                 
\bibitem{lakshmi} S. Lakshmi, S. Roche, and G. Cuniberti, Phys. Rev. B \textbf{80}, 193404 (2009).            
     
\bibitem{LTP} D. O. Oriekhov, E. V. Gorbar, and V. P. Gusynin, Low Temp. Phys. \textbf{44} 1313 (2018).

\bibitem{onsagar1} L. Onsager, Phys. Rev. \textbf{37}, 405 (1931)

\bibitem{onsagar2} L. Onsager, Phys. Rev. \textbf{38}, 2265 (1931)

\bibitem{dutta} S. Datta, Electronic Transport in Mesoscopic Systems
(Cambridge University Press, Cambridge, England, 1995).

\bibitem{lv} S. H. Lv and Y. X. Li, J. Appl. Phys. \textbf{112}, 053701 (2012).

\bibitem{groth} C. W. Groth, M. Wimmer, A. R. Akhmerov, and X. Waintal, New J. Phys. \textbf{16}, 063065 (2014).

\bibitem{para_th} P. Dutta, A. Saha and A. M. Jayannavar, Phys. Re. B \textbf{96}, 115404 (2017).

\bibitem{Ferrer} J. Ferrer et. al., New J. Phys. \textbf{16}, 093029 (2014).

\bibitem{hatef} H. Sadeghi, S. Sangtarash and Colin J. Lambert, Beilstein J. Nanotechnol. \textbf{6}, 1176 (2015).

\bibitem{sancho} M. L. Sancho et al., J. Phys. F \textbf{14}, 1205 (1984).

\bibitem{para} P. Dutta, S. K. Maiti, and S. Karmakar, J. Appl. Phys. \textbf{114}, 034306 (2013).

\bibitem{silicene} Kh. Shakouri, H. Simchi, M. Esmaeilzadeh,
H. Mazidabadi, and F. M. Peeters, Phys. Rev. B \textbf{92}, 035413 (2015).

\bibitem{mos2} F. Khoeini, Kh. Shakouri, and F. M. Peeters, Phys. Rev. B \textbf{94}, 125412 (2016).

\bibitem{njp_ezawa} M. Ezawa, New J. Phys. \textbf{16}, 115004 (2014).

\bibitem{coulomb} L. Yang et al., Phys. Rev. Lett \textbf{99}, 186801 (2007).

\bibitem{DFT} V. Valeria et al., Phys. Rev. B \textbf{87}, 115117 (2013).

\bibitem{mate} Mate Vigh et al., Phys. Rev. B \textbf{88} 161413 (R) (2013).

\bibitem{defects1} Y. Kobayashi, K.-I. Fukui, T. Enoki, and K. Kusakabe, Phys. Rev. \textbf{73}, 125415 (2006).

\bibitem{defects2} Y. Niimi, T. Matsui, H. Kambara, K. Tagami, M. Tsukada, and H. Fukuyama, Phys. Rev. \textbf{B} 73, 085421 (2006).

\bibitem{peeters} R. N. Costa Filho, G. A. Farias, and F. M. Peeters, Phys. Rev. B \textbf{76}, 193409 (2007)

\bibitem{white} D. Gunlycke and C. T. White, Phys. Rev. Lett. \textbf{106}, 136806 (2011)

\end{thebibliography}
\end{document}